\documentclass{article}
\usepackage{arxiv}

\usepackage[utf8]{inputenc} 
\usepackage[T1]{fontenc}    
\usepackage[hidelinks]{hyperref}       
\usepackage{url}            
\usepackage[version=4]{mhchem}
\usepackage{amsfonts}       
\usepackage{mathtools}
\usepackage{nicefrac}       
\usepackage{microtype}      
\usepackage[square,numbers,sort&compress]{natbib}
\usepackage{doi}

\usepackage{graphicx}%
\usepackage{multirow}%
\usepackage{amsmath,amssymb,amsfonts}%
\usepackage{amsthm}%
\usepackage{mathrsfs}%
\usepackage{rotating}
\usepackage[title]{appendix}%
\usepackage{xcolor}%
\usepackage{textcomp}%
\usepackage{manyfoot}%
\usepackage{booktabs}%
\usepackage{tabularray}
\usepackage{caption}
\usepackage{algorithm}%
\usepackage{algorithmicx}%
\usepackage{algpseudocode}%
\usepackage{listings}%
\usepackage{soul}
\usepackage{subfig}
\usepackage{graphicx}
\usepackage{amssymb}
\usepackage{pifont}
\usepackage{subfloat}

\usepackage{tabularx}
\usepackage{tabu}
\usepackage{tikz}
\definecolor{navy}{RGB}{0,114,189}
\definecolor{red}{RGB}{230, 77, 26}
\definecolor{yellow}{RGB}{235, 165, 80}

\usepackage{subcaption}
\usepackage{siunitx}

\newcommand{\Rey}{{\mbox{\textit{Re}}}} 
\definecolor{DarkRed}{rgb}{0.65,0.00,0.05}
\definecolor{DarkGreen}{rgb}{0.05,0.65,0.05}
\definecolor{DarkOrange}{RGB}{255,77,1}
\definecolor{greenFig}{RGB}{0,160,60}

\usepackage[normalem]{ulem}

\title{Impact of Structure-Preserving Discretizations on Compressible Wall-Bounded Turbulence of Thermally Perfect Gases}%

\author{ \href{https://orcid.org/0009-0009-5872-702X}{\includegraphics[scale=0.06]{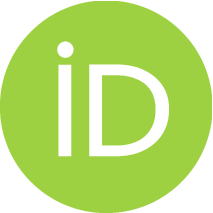}\hspace{1mm} Alessandro {Aiello}}\\
	Dipartimento di Ingegneria Industriale\\
	Universit\`a di Napoli ``Federico II''\\
	Napoli, Italy \\
	\texttt{alessandro.aiello@unina.it} \\
	\And
    \href{https://orcid.org/0000-0003-0274-5260}{\includegraphics[scale=0.06]{orcid.eps}\hspace{1mm} Andrea {Palumbo}}\\
    Dipartimento di Ingegneria Meccanica e Aerospaziale\\
    Sapienza Università di Roma\\
	Roma, Italy \\
	\texttt{andrea.palumbo@uniroma1.it} \\
	\And
    \href{https://orcid.org/0000-0002-6518-3114}{\includegraphics[scale=0.06]{orcid.eps}\hspace{1mm} Carlo {De~Michele}}\\
	Mathematics Area\\
	Grans Sasso Science Institute\\
	L'Aquila, Italy \\
	\texttt{carlo.demichele@gssi.it} \\
	\And
	\href{https://orcid.org/0000-0003-4943-9551}{\includegraphics[scale=0.06]{orcid.eps}\hspace{1mm}Gennaro Coppola} \\
	Dipartimento di Ingegneria Industriale\\
	Universit\`a di Napoli ``Federico II''\\
	Napoli, Italy \\
	\texttt{gcoppola@unina.it} \\
}

\renewcommand{\shorttitle}{Impact of Structure-Preserving Discretizations on Compressible Turbulence of TP Gases}%

\hypersetup{
pdftitle={Impact of Structure-Preserving Discretizations on Compressible Wall-Bounded Turbulence of Thermally Perfect Gases},
pdfsubject={physics.flu-dyn},
pdfauthor={Alessandro Aiello, Andrea Palumbo, Carlo De~Michele, Gennaro Coppola}
pdfkeywords={compressible turbulence, thermally perfect gas model, structure-preserving discretizations, entropy-conservative methods, direct numerical simulation}
}

\begin{document}
\maketitle

\begin{abstract}
Direct numerical simulations of compressible turbulent channel flow at supersonic and hypersonic Mach numbers are performed using a thermally perfect gas model for CO$_2$. The objective is to assess the role of structure-preserving discretizations of the convective terms in high-enthalpy regimes, with particular emphasis on entropy conservation, kinetic-energy preservation, and consistency with the thermodynamic closure.

The comparative analysis of various formulations examines their impact on robustness, thermodynamic fluctuations, and turbulence statistics across a range of Mach numbers. Differences among formulations are found to originate primarily in the treatment of thermodynamic variables and progressively influence the dynamical fields as compressibility effects intensify. In particular, the coupling between entropy consistency and pressure discretization is shown to affect Reynolds stresses and mean flow properties in high-speed regimes.

Overall, the results indicate that consistency between the numerical formulation and the thermodynamic model contributes significantly to the reliable simulation of high-enthalpy compressible turbulence. The study systematically assesses entropy-conservative discretizations for thermally perfect gases in wall-bounded flows and examines their impact on thermodynamic-dynamic coupling at high Mach numbers. 
\end{abstract}

\section{Introduction} 
Direct numerical simulation (DNS) has long been a central tool for the study of wall-bounded turbulent flows, as it provides fully resolved datasets that enable detailed investigation of turbulence structure, scaling behavior, and transport mechanisms without reliance on turbulence models.
Among canonical configurations, pressure-driven turbulent channel flow occupies a prominent role owing to its geometric simplicity, statistical homogeneity in the streamwise and spanwise directions, and relevance to both fundamental turbulence research and engineering applications.
Seminal incompressible DNS studies by \citet{Kim_Moin_Moser_1987} and \citet{Moser_Kim_Mansour_1999} established quantitative reference data for mean velocity profiles, turbulence intensities, and near-wall dynamics.
Subsequent simulations extended these results to progressively higher friction Reynolds numbers, providing benchmark datasets that continue to underpin research in turbulence theory~\citep{Hoyas2006, bernardini2014, Lee_Moser_2015,Hoyas2022}.

The extension of DNS to compressible wall-bounded flows enabled systematic investigation of the effects of density variations, compressibility, and thermodynamic fluctuations on turbulence dynamics. Supersonic channel and boundary-layer flows have therefore been widely adopted as canonical configurations for high-speed wall turbulence, in which momentum and thermal fields are strongly coupled. Early DNS of compressible turbulent channel flow by \citet{Coleman_Kim_Moser_1995} and \citet{Huang_Coleman_Bradshaw_1995}, conducted for ideal-gas flows between isothermal walls at bulk Mach numbers up to $M_b \approx 3$, provided evidence that many features of wall turbulence remain qualitatively similar to their incompressible counterparts. These studies supported Morkovin’s hypothesis~\citep{Morkovin1962}, according to which, at sufficiently high Reynolds number and moderate Mach number, compressibility effects influence wall turbulence primarily through spatial variations of mean density and transport properties, while density fluctuations play a secondary role.

Building on these observations, subsequent investigations introduced semi-local scaling concepts, in which local values of density and viscosity are used to define wall units and Reynolds numbers~\cite{Huang_Coleman_Bradshaw_1995, Trettel2016, Patel2017}. Such scalings were shown to improve the collapse of turbulent statistics relative to the classical van Driest transformation~\cite{VanDriest_1951}.
DNS databases spanning a wide range of Mach and Reynolds numbers have demonstrated that, under appropriate conditions and scaling, high-speed flow in channel shares some common traits with their incompressible counterpart~\citep{Foysi_2004, Morinishi_2004, Tamano_2006, Modesti2016}. These results have often been interpreted as indicating that compressibility acts in a largely indirect manner in canonical wall-bounded flows, modulating turbulence primarily through mean thermodynamic and transport-property variations. More recent studies suggest that intrinsic compressibility effects can be dynamically significant, modulating the near-wall dynamics of turbulence~\cite{hasan2023,hasan2025}. In this context, cold wall conditions appear to further modify the near-wall flow dynamics by affecting vortex topology and turbulent energy production
\cite{yu2019, yu2021}, especially for hypersonic flow.

From a theoretical perspective, compressible flows may be decomposed into vortical, acoustic, and entropy modes following the linear analysis of \citet{Chu1958}. While this decomposition does not constitute a derivation of Morkovin’s hypothesis, it provides a useful interpretative framework. For wall-bounded turbulent flows, extensive experimental and numerical evidence indicates that turbulence production is dominated by vortical motions interacting with mean shear, whereas acoustic modes contribute weakly in the absence of strong shock–turbulence interaction~\citep{Spina1994}. Entropy modes are linearly passive in the sense that they do not directly generate vorticity or turbulence kinetic energy; however, they influence the flow indirectly through their impact on density, temperature, and transport properties, thereby modulating dissipation, diffusion, and heat transfer~\cite{Wang2019}. The latter feature highlights the need of solving all those components using proper numerical methods, as compressibility-related effects increase dramatically with the bulk flow Mach number.

These considerations become increasingly important when moving beyond the assumption of a Calorically Perfect (CP) gas. While the CP model is adequate for moderate temperature variations, many high-speed and high-enthalpy flows exhibit temperature-dependent specific heats and transport properties~\cite{Duan2011}. In such Thermally Perfect (TP) or more general gas models, entropy variations are no longer trivially related to temperature, and the resulting changes in viscosity, thermal conductivity, and Prandtl number can be significant~\cite{Sciacovelli2017}.

From a numerical standpoint, DNS of compressible turbulent flows at high Reynolds numbers poses substantial challenges. Standard discretizations of the nonlinear convective terms are prone to numerical instabilities, even in smooth, shock-free flows, due to the amplification of discretization errors across a wide range of interacting scales. Stabilization strategies based on artificial dissipation or filtering can prevent catastrophic breakdown, but often at the expense of excessive numerical dissipation that distorts energy transfer and turbulence statistics. These limitations have motivated the development of structure-preserving discretizations designed to respect key invariants and balance laws of the governing equations at the discrete level~\cite{Veldman_SIAMRev_2021}.

In incompressible flows, the pioneering work of \citet{Arakawa_JCP_1966} and \citet{Lilly_JCP_1965} demonstrated that suitable discretizations of the nonlinear terms can exactly preserve quadratic invariants such as kinetic energy and/or enstrophy. In the compressible setting, preservation of kinetic energy by the discrete convective operator was also identified as an important ingredient for robust simulations~\citep{Feiereisen_1981}, and Kinetic-Energy-Preserving (KEP) formulations were shown to significantly enhance numerical stability without resorting to numerical dissipation~\cite{Pirozzoli_JCP_2010,Coppola_JCP_2019,Coppola_AMR_2019}.
In parallel, entropy has emerged as a fundamental quantity governing both mathematical admissibility and physical consistency of solutions of the compressible Euler and Navier--Stokes equations. The entropy framework introduced by \citet{Tadmor_MC_1987} established algebraic conditions for discrete entropy conservation and entropy stability, leading to the development of entropy-conservative and entropy-stable schemes for compressible flows.
These developments have led to the modern view that robust and predictive simulations of compressible turbulent flows require discretizations that respect both kinetic energy and entropy principles at the discrete level~\cite{Chandrashekar_CCP_2013,Ranocha_JSC_2018,Slotnick_NASA_2014}.

While entropy-based discretizations have traditionally been assessed using idealized test cases, their application to wall-bounded turbulent flows has only recently begun to be explored. Existing studies have primarily focused on calorically perfect gas models and have demonstrated the feasibility of entropy-aware formulations in turbulent channel flows and related configurations~\cite{Parsani2021,Schwarz2025,Alberti_2025}.
Extending such approaches to thermally perfect gases is nontrivial, as the algebraic structure underlying classical entropy formulations is altered when specific heats depend on temperature. Recent developments have addressed this challenge by deriving entropy-conservative discretizations for general equations of state and thermally perfect gases~\citep{Aiello_JCP_2025, Gouasmi_CMAME_2020,Peyvan_JCP_2023,Oblapenko_C&F_2025, Aiello_ArXiv_2025}.

In this context, entropy-conservative discretizations represent a physically consistent and promising framework for DNS of supersonic turbulent channel flows with temperature-dependent properties. By preserving the correct entropy balance in smooth regions, such schemes may be particularly well suited to capturing the indirect coupling between turbulence, entropy fluctuations, and transport-property variations that characterizes high-speed wall-bounded flows.
The present work builds upon these developments and presents direct numerical simulations of compressible turbulent channel flow of a thermally perfect gas using entropy-conservative schemes, with the aim of assessing the impact of both thermodynamic modeling and numerical discretization on wall-bounded compressible turbulence.

The remainder of the paper is organized as follows. In Section~\ref{sec:model}, the governing equations, the thermodynamic model, and the channel flow configuration are presented, including non-dimensional parameters, averaging procedures, numerical domain, and boundary conditions. Section~\ref{sec:mathematical_formulation} details the numerical methods and spatial discretizations employed, with particular emphasis on entropy-conservative formulations. The results of the turbulent channel flow simulations are presented in Section~\ref{sec:results}, whereby the effect of the numerical method is shown on the main flow velocity and thermodynamic variables statistics. Finally, the main findings and conclusions are summarized in Section~\ref{sec:conclusions}.
\section{Governing equations and fluid model}\label{sec:model}

The three-dimensional Navier--Stokes equations for mass, momentum and total energy are solved for a viscous, heat-conducting fluid in a Cartesian coordinate system $x_i,\,i=1,2,3$:
\begin{align}
    \frac{\partial \rho}{\partial t} + \frac{\partial \rho u_i}{\partial x_i} & = 0\label{eq:rho}\\
    \frac{\partial \rho u_j}{\partial t} + \frac{\partial \rho u_i u_j}{\partial x_i} & = - \frac{\partial p}{\partial x_j} + \frac{\partial \tau_{ij}}{\partial x_i} + f_i\delta_{i1} \label{eq:rhou}\\
    \frac{\partial \rho E}{\partial t} + \frac{\partial \rho u_i H}{\partial x_i}  &= \frac{\partial \tau_{ij}u_j}{\partial x_i} - \frac{\partial q_i}{\partial x_i} + f_1u_1\label{eq:rhoE}
\end{align}
with $\rho$ and $p$ being the density and the thermodynamic pressure, respectively, $u_i$ the Cartesian components of the velocity field and $H=k+e+p/\rho = E + p/\rho$ the total enthalpy, expressed as the sum of kinetic energy $k=u_iu_i/2$, internal energy $e$ and pressure-volume work. Total energy ($E$) and kinetic and internal energy are per unit mass. Finally, $\tau_{ij}$ is the dissipative part of the stress tensor $\sigma_{ij}=-p\delta_{ij} + \tau_{ij}$, and $q_j$ is the heat flux. These quantities are modeled as after the Newtonian fluid model and the Fourier's law, respectively
\begin{equation*}
    \tau_{ij} = \mu\left(\frac{\partial u_j}{\partial x_i} + \frac{\partial u_i}{\partial x_j}\right) - \frac{2}{3}\mu\frac{\partial u_k}{\partial x_k}\delta_{ij}\,,\qquad q_j = -\lambda\frac{\partial T}{\partial x_j},
\end{equation*}
where $\mu$ is the dynamic viscosity and $\lambda$ the thermal conductivity of the gas, and $T$ is the temperature. The forcing term $f_i$ is included in the momentum and total energy equations to enforce a constant mass flow rate through the channel.
System \eqref{eq:rho}--\eqref{eq:rhoE} is closed upon specification of an equation of state and the fundamental thermodynamic relation, respectively in the form $p=p(e,\rho)$ and $e=e(s,\rho)$, with $s$ the specific entropy per unit mass.

In this work, we adopt the thermally perfect gas model to close the Navier--Stokes equations, where pressure is given by the relation $p=\rho RT$, with $R$ being the specific gas constant.
Internal energy and entropy are linked to temperature through the specific heat capacities at constant pressure and volume (respectively indicated as $c_p(T)$ and $c_v(T)$), following the relations
\begin{equation*}
    e = e_{\mathrm{ref}} + \int_{T_{\mathrm{ref}}}^Tc_v(T')\,\mathrm{d} T'\,,\qquad s = s_{\mathrm{ref}} + \int_{T_{\mathrm{ref}}}^T\frac{c_v(T')}{T'}\,\mathrm{d}T'- R\log{\frac{\rho}{\rho_{\mathrm{ref}}}}.
\end{equation*}

Their use requires the knowledge of one of the specific heat capacities $c_v(T)$ or $c_p(T)$, since Mayer's relation $c_p=R+c_v$ holds.
In our case, it is obtained from the 7-coefficient polynomial fittings reported in~\cite{NASA}; different models may be selected for specific applications, without loss of generality for the numerical formulation presented hereafter.
The thermally perfect gas model is widely used in applications where high temperatures or wide temperature ranges occur, as for high-enthalpy and/or highly compressible flow cases.

In the present work, we are presenting direct numerical simulations of turbulent compressible channel flows with carbon dioxide (CO$_2$) up to hypersonic regime in high-enthalpy conditions. Analyzing CO$_2$ airflow under a non-reactive approximation is appropriate for some space engineering applications, including the ending phase of the atmospheric landing on Mars, whose atmosphere is predominantly composed by carbon dioxide (and small amounts of nitrogen and argon)~\cite{placco2025}. In fact, at moderate to high temperatures as those considered in this study, it is often sufficient to model carbon dioxide as a thermally perfect gas, given its non-polar molecular structure and the resulting negligible contribution of van der Waals forces.

Equations \eqref{eq:rho}--\eqref{eq:rhoE} are projected on a Cartesian reference frame, whereby the $x$ axis runs along the streamwise direction, $y$ is the wall-normal direction, and $z$ the spanwise one. The flow is considered to be statistically homogeneous along the streamwise and spanwise directions, hence periodic boundary conditions are enforced at the $x$- and $z$-normal boundaries. No-slip conditions are imposed on the channel walls $y=0$ and $y=2h$ (where $h$ is the channel half-width), whose temperature $T_w$, corresponding to the dimensional reference value of $T_{\mathrm{ref}}=298.15$ K, is fixed during the computation. The chosen numerical set-up yields a constant bulk velocity $U_b$ and a variable bulk temperature $T_b$, which evolves in time until a statistically steady state is reached. An important non-dimensional parameter characterizing the flow under study is the bulk Reynolds number, which is defined as $\Rey_b=\rho_bU_bh/\mu_w$, where dynamic viscosity is modeled according to the Sutherland's law, i.e.
\begin{equation*}
    \frac{\mu(T)}{\mu(T_{\mathrm{ref}})}=\left(\frac{T}{T_{\mathrm{ref}}}\right)^{3/2}\frac{(1+C/T_{\mathrm{ref}})}{(T/T_{\mathrm{ref}}+C/T_{\mathrm{ref}})},
\end{equation*}
with $C=\SI{220.2}{\kelvin}$ for CO$_2$.
The Mach number can be defined either referring to wall values, $M_b^w=U_b/\sqrt{\gamma_wRT_w}$, or to bulk conditions, $M_b=U_b/\sqrt{\gamma_bRT_b}$, with $\gamma(T)=c_p(T)/c_v(T)$ the ratio of specific heat capacities.
In this study,
the choice of the input Mach number uniquely determines the wall heat flux, which is fully determined by the imposed Mach number through the balance between viscous dissipation and thermal diffusion.

The numerical data-sets obtained in this work are processed to obtain turbulent flow statistics. For a generic quantity $\phi$, its Reynolds decomposition reads $\phi = \langle \phi \rangle + \phi'$, where $\langle \phi \rangle$ is the time-average and $\phi'$ the corresponding fluctuation. Favre averaging is also considered in this work, viz. $\phi = \tilde{\phi} + \phi''$, where
$\tilde{\phi}=\langle \rho \phi\rangle/\langle\rho\rangle$. For all quantities, streamwise and spanwise averaging is also performed, as after flow homogeneity along these directions. In the following, we are dropping the brackets for wall and centerline conditions for the sake of simplifying the notation.

It is useful to introduce inner-scaled quantities to characterize the flow. We adopt the common definition for friction Reynolds number as $\Rey_\tau=u_\tau \rho_w h/\mu_w$, with $u_\tau=(\tau_w/\rho_w)^{1/2}$ and $\tau_w=\langle \mu_w(\mathrm{d}u/\mathrm{d}y)_w \rangle$ being the friction velocity and the time-averaged value of wall friction. Accordingly, viscous units are introduced as $x_i^+=x_i/\delta_\nu $, with $\delta_\nu=\mu_w/(\rho_w u_\tau)$ the viscous length-scale at the wall, whereas heat transfer will be measured through the nondimensional heat flux coefficient $B_q=q_w/(\rho_wc_{p_w}u_\tau T_w)$. We also use the wall friction Mach number $M_\tau=u_\tau/\sqrt{\gamma_wRT_w}$ as a measure of near-wall compressibility effects~\cite{hasan2023} and the transformed friction Reynolds number~\cite{Trettel2016} $\Rey_\tau^*=\rho_c(\tau_w/\rho_c)^{1/2}h/\mu_c$, where the subscript “$c$" indicates centerline conditions at $y=h$, to determine an equivalent incompressible state. The wall-normal direction will be reported in semi-local scaling as well, that is $y^*=y(\tau_w\langle{\rho}\rangle)^{1/2}/\langle \mu \rangle$.
\section{Numerical Methodology}\label{sec:mathematical_formulation}
The present work focuses on obtaining novel simulations of high-enthalpy channel flows of a compound gas in supersonic/hypersonic regimes with a refined numerical treatment of the convective terms of the discretized equations as to induce additional structural properties. With respect to the main system \eqref{eq:rho}--\eqref{eq:rhoE}, the Eulerian part of each equation can be written in the general form
\begin{equation*}
    \frac{\partial \rho \phi}{\partial t} = -\mathcal{C}_{\rho\phi} - \mathcal{P}_{\rho\phi}:=-\mathcal{R}_{\rho\phi}
\end{equation*}
with $\phi=1,u_i,E$ for Eqs.~\eqref{eq:rho}, \eqref{eq:rhou} and \eqref{eq:rhoE}, respectively.  $C_{\rho\phi}$ and $\mathcal{P}_{\rho\phi}$ are the proper advection and pressure terms, respectively, with $C_{\rho\phi}$ having the divergence structure $\partial(\rho u_j \phi)/\partial x_j$.
Compressible Euler equations represent a system of conservation laws, which means that, on domains with vanishing boundary effects,
\begin{equation*}
    \frac{\mathrm{d}}{\mathrm{d} t}\int_{\Omega}\rho \phi\,\mathrm{d}V=-\int_\Omega\mathcal{R}_{\rho\phi}\,\mathrm{d}V=0
\end{equation*}
In the continuous framework, the same applies for other quantities, such as specific thermodynamic entropy, $\rho s$, and for the convective term of specific kinetic energy $\mathcal{C}_{\rho k}$. In fact, for compressible, nonviscous and adiabatic flows, we have
\begin{align}
\frac{\partial \rho k}{\partial t} = -\frac{\partial \rho u_i k}{\partial x_i} -u_i\frac{\partial p}{\partial x_i} \;\longrightarrow\,\int_\Omega \mathcal{C}_{\rho k}\,\mathrm{d}V = 0\label{eq:rhok}\\
    \frac{\partial \rho s}{\partial t} = -\frac{\partial \rho u_i s}{\partial x_i} \;\longrightarrow\,\int_\Omega \mathcal{C}_{\rho s}\,\mathrm{d}V = 0.\label{eq:rhos}
\end{align}
Both kinetic energy and entropy are examples of what we label as \emph{induced} variables, since their evolution is driven by the discretization of the main system of equations, and, in general, in the discrete framework, they are not consistent with the continuous models \eqref{eq:rhok}--\eqref{eq:rhos} unless some \emph{ad hoc} treatment is employed. Numerical methods that are able to discretely reproduce conditions \eqref{eq:rhok}--\eqref{eq:rhos} are termed as \emph{(globally) kinetic-energy preserving} (KEP) and \emph{(globally) entropy-conservative} (EC), respectively, and their fulfillment has been proved to significantly enhance fidelity and robustness of fluid-flow computations. On the other hand, \emph{local} conservation
states the possibility of recasting each convective term as a difference of numerical fluxes, that is $\mathcal{C}_{\rho\phi}|^{i+1/2}=\delta^-\mathcal{F}_{\rho\phi}^{\,c,\,i+1/2}/\Delta x^i$, with $\delta^-a^i=a^{i}-a^{i-1}$ being the backward finite-difference operator and the superscript “c'' indicating the convective part of the numerical flux~\cite{Jameson_JSC_2008,DeMichele_C&F_2023}. Note that, to avoid confusion with the indices used to denote Cartesian components, we are denoting nodal values as superscripts.
Local conservation implies global conservation by virtue of the telescoping property, and the inverse implication also holds true for a wide class of finite-difference discretizations~\cite{Coppola_JCP_2023}.
Considering collocated meshes, numerical fluxes are constructed via interpolation between adjacent cell-center values---for the second-order case---, that is $\mathcal{F}_{\rho\phi}^{i+1/2}=\widehat{(\rho,u_i,\phi)}^{i,i+1}$. For pressure terms $\mathcal{P}_{\rho\phi}^{i}$ the construction is similar~\cite{Aiello_JCP_2025}. The high-order extension has been obtained multiple times (e.g.~\cite{LeFloch_SIAMJNA_2002,Pirozzoli_JCP_2010,Ranocha_JSC_2018,DeMichele_JCP_2023}) and it is constructed by considering linear combinations of the two-point fluxes on wider stencils.
Finally, for the second-order case, we usually drop the apex $i + 1/2$ for notation purposes.

Numerical analysis of the Navier--Stokes system, in our framework, focuses on proper \emph{spatial} discretization aimed at achieving both exact KEP and EC properties. For simplicity, and without loss of generality, we consider the one-dimensional case on a uniform mesh with periodic boundary conditions. Generalization to multi-dimensional cases is easily carried out by directions and non-uniform grids are treated by considering local spacing $\Delta x^{i,j,k}$. Finally, when non-periodic boundary conditions are considered, such constructions still apply at interior points, and a special treatment of boundary closure schemes has to be considered. Assuming sufficiently small time-integration errors, semi-discretization of the Eulerian part of system \eqref{eq:rho}--\eqref{eq:rhoE} with central formulas yields\
\begin{equation}
    \frac{\mathrm{d}}{\mathrm{d} t}\begin{Bmatrix}
        \rho\\
        \rho u\\
        \rho E
    \end{Bmatrix} = -\frac{1}{\Delta x}\delta^-\begin{Bmatrix}
        \mathcal{F}_{\rho}^c\\
        \mathcal{F}_{\rho u}^c\\
        \mathcal{F}_{\rho E} ^c
    \end{Bmatrix}-\begin{Bmatrix}
        0\\
        \mathcal{P}_{\rho u}\\
        \mathcal{P}_{\rho E}
    \end{Bmatrix}.
\end{equation}
It can be shown that the KEP property obtained by discretely enforcing Eq.~\eqref{eq:rhok} induces a constraint on the relation between mass and momentum numerical fluxes, which for second-order schemes reduces to~\cite{Jameson_JSC_2008b,Veldman_JCP_2019} $ \mathcal{F}_{\rho u}^c = \mathcal{F}_\rho \overline{u}$, leaving as arbitrary (as long as consistent with the continuous counterpart) the choice of the mass flux and of the pressure term in momentum equation.
We consider here the simple formulation given by
$\mathcal{F}_\rho = \hat{\rho}\,\overline{u},\,\mathcal{F}_{\rho u}=\mathcal{F}_{\rho}\overline{u} + \overline{p}$, with $\hat{\rho}$ a consistent approximation of density at $x_{i+1/2}$ and $\overline{\phi}=(\phi^i+\phi^{i+1})/2$ the arithmetic average.
Such choice induces both numerical fluxes and pressure terms for kinetic energy as $\mathcal{F}_{\rho k}^c  =\mathcal{F}_\rho u^iu^{i+1}/2$ and $\mathcal{P}_{\rho k} = u^i\delta^-\overline{p}/\Delta x$~\cite{DeMichele_C&F_2023}. By definition, the existence of numerical fluxes is the aforementioned \emph{local} conservation, which implies conservation of $\rho k$, by convection, in the integral sense. Thus, total energy can be discretized as the sum of its kinetic-energy and internal-energy contributions, that is $\mathcal{F}_{\rho E}=\mathcal{F}_{\rho}u^iu^{i+1}/2+ \mathcal{F}_{\rho e}^c+\overline{\overline{(u,p)}}$, with $\overline{\overline{(u,p)}}=(u^ip^{i+1}+u^{i+1}p^i)/2$ the product-mean.

Finally, the general form of the KEP discretization we consider is given by the set of fluxes
\begin{equation}\label{eq:KEP}
\mathcal{F}_\rho=\hat{\rho}\,\overline{u}\,,\qquad \mathcal{F}_{\rho u}=\mathcal{F}_\rho \overline{u}\,+\overline{p},\qquad \mathcal{F}_{\rho E} = \frac{\mathcal{F}_\rho u^iu^{i+1}}{2}+\mathcal{F}_{\rho e}^c + \overline{\overline{(u,p)}}
\end{equation}
Conditions \eqref{eq:KEP} on the mass and momentum equations are seen to provide fulfillment of the KEP property independently on the discretization of the internal-energy convective contribution $\mathcal{F}_{\rho e}^c$, and in turn, on the numerical treatment of the energy equation.
In fact, $\mathcal{F}_{\rho e}^c$ and the interpolation used for density, $\hat{\rho}$, represent additional degrees of freedom that can be exploited to enforce additional structural properties. This is the case, among the others, of entropy conservation.
Proper discretizations of the energy equation aimed at enforcing additional thermodynamic properties require consistency with the chosen gas models and, in general, with the equation of state. It is noteworthy to remark that KEP discretizations will always be used in this work, and EC properties has to be seen as an additional feature to the latter, both in the ideal-gas case and in the non-ideal one.

For single-component CP gases, formulations of KEP and EC---the latter satisfied within different levels of approximations---fluxes have been developed~\cite{Chandrashekar_CCP_2013,Ranocha_JSC_2018,DeMichele_JCP_2025} and are implemented in many high-performance solvers~\cite{Bernardini_CPC_2023,ranocha2022adaptive}. In practical applications involving non-CP gases, these formulations are often employed nonetheless, at the expense of their original structural properties, due to the lack of established alternatives. The kinetic energy- and entropy-preserving formulation (KEEP) of \citet{Kuya_JCP_2018} reproduces discrete EC approximately by means of the arithmetic average on density and the so-called Kennedy--Gruber--Pirozzoli (KGP, \cite{Kennedy_JCP_2008,Pirozzoli_JCP_2010,Coppola_JCP_2019}) split on internal energy. Its performances have been afterwards enhanced making use of Taylor-series expansions to reduce the entropy production error down to machine precision~\cite{Tamaki_JCP_2022,Kawai_JCP_2025}.
Ranocha proposed an EC and KEP formulation which is also Pressure-Equilibrium Preserving (PEP), that allows to reproduce traveling density waves at constant velocity and pressure and significantly improve stability of numerical simulations~\cite{Ranocha_JSC_2018,Ranocha_CAMC_2021}.
The latter makes use of the logarithmic mean $\overline{\phi}^{\mathrm{log}}=(\phi^{i+1}-\phi^i)/(\log{\phi}^{i+1}-\log\phi^i)$, where the potential singularity is treated by substituting the fix proposed by \citet{Ismail_JCP_2009} up to a specific tolerance.
Both singularity and the intrinsic computational cost of the logarithmic mean have been addressed by \cite{DeMichele_JCP_2023} who derived and Asymptotically Entropy-Conservative (AEC) scheme, based on the Taylor-series expansion of the logarithmic terms, which is KEP and asymptotically achieves entropy conservation, while also being PEP~\cite{DeMichele_JCP_2024}.

As the analysis shifts to non-CP gas conditions---either thermally perfect or real-gas models---the aforementioned formulations no longer retain EC properties, although they still enforce the KEP property, which depends exclusively on the discretization of mass and momentum equations. For single-component gases governed by arbitrary equations of state, an EC spatial discretization has been recently derived in~\cite{Aiello_JCP_2025}, by imposing discrete consistency with the Gibbs' relation in terms of volumetric quantities $\partial \rho e/\partial t=T\partial \rho s/\partial t + g\partial \rho/\partial t$, with $g=e-Ts+p/\rho$ the Gibbs' free energy. Such formulation, as for the ideal-gas case, presents a singularity for uniform temperature distributions. Upon specification of the specific gas model---the thermally perfect one, in this work---singularity can be treated with either the fix by Ismail and Roe or the asymptotic expansions~\cite{Aiello_ArXiv_2025} following the AEC formulation for the CP case. This discretization, that we will label EC-TP,
will be used in Section~\ref{sec:results} in its logarithmic form for comparison purposes with other existing methods that do not rely on the asymptotic approximations~\cite{Gouasmi_CMAME_2020}. Finally, the EC-TP discretization reads~\cite{Aiello_ArXiv_2025}
\begin{align*}
    \mathcal{F}_\rho &= \overline{\rho}^{\mathrm{log}}\overline{u},\\
    \mathcal{F}_{\rho u} &= \mathcal{F}_{\rho}\overline{u} + \overline{p},\\
    \mathcal{F}_{\rho E} &= \mathcal{F}_{\rho}\frac{u^iu^{i+1}}{2} + \mathcal{F}_{\rho}\bigg[ \epsilon_{\mathrm{ref}} +
    c_{-1}\left(1-\dfrac{\overline{1/T}}{\overline{1/T}^{\mathrm{log}}} + \overline{\log T}\right)+\\&+c_0\dfrac{1}{\overline{1/T}^{\mathrm{log}}} + \sum_{m=-\ell}^{-2}\frac{c_m}{m+1}\widetilde{\widetilde{T}}^m +\sum_{m=1}^{r}\frac{c_m}{m+1}\overline{\overline{T}}^m\bigg] + \overline{\overline{(u,p)}},
\end{align*}
where $c_m,\,m\in[-\ell,r]$ are the coefficients belonging to (one) polynomial fitting for heat capacities as functions of temperature, that is $c_v(T)=\sum_{m=-\ell}^rc_m(T)$ $\widetilde{\widetilde{T}}^m$~\cite{NASA,NASA_CHASE}, while $\epsilon_{\mathrm{ref}}$ is related to the reference conditions (also tabled in~\cite{NASA}). Finally, $\overline{\overline{T}}^{\,m} = m^{-1}{\sum_{\nu=1}^{m}T_{i+1}^{m-\nu+1}T_i^{\nu}}$,for $m\geq 1$, and $\widetilde{\widetilde{T}}^{\,m} = {|m|}^{-1}{\sum_{\nu=m+1}^{0}T_{i+1}^{m-\nu+1}T_i^{\nu}}$,\ for $m\leq -2$, can be seen as generalizations of existing averages. A comprehensive comparative overview of the main characteristics and properties of the tested and analyzed formulations is provided in Table~\ref{tab:schemes}.
\begin{sidewaystable}
    \centering
    {
    \begin{tabular}{c c c c c c c c}
    \toprule
    \addlinespace
        \textbf{Scheme} &  \textbf{KEP} & \textbf{EC} & $\boldsymbol{\mathcal{F}_\rho}$ & $\boldsymbol{\mathcal{F}_{\rho u}^c}$& $\boldsymbol{{\mathcal{P}}_{\rho u}}$& $\boldsymbol{\mathcal{F}_{\rho e}^c}$ &$\boldsymbol{\mathcal{P}_{\rho E}}$\vspace{0.2cm}\\
        \midrule\\
        EC-TP~\cite{Aiello_ArXiv_2025}  & \ding{51} &\ding{51}&$\overline{\rho}^{\mathrm{log}}\overline{u}$&$\mathcal{F}_\rho\overline{u}$&$\overline{p}$&$\mathcal{F}_{\rho}\left[\epsilon_{\mathrm{ref}} +
    c_{-1}\left(1-\dfrac{\overline{1/T}}{\overline{1/T}^{\mathrm{log}}} + \overline{\log T}\right)+\dfrac{c_0}{\overline{1/T}^{\mathrm{log}}} + \xi
    \right]$&$\overline{\overline{(u,p)}}$\\

    \vspace{0.1cm}\\
    AEC$^{(N)}$-TP \cite{Aiello_ArXiv_2025} & \ding{51} &{\ding{109}}&$\dfrac{\overline{\rho}\,\overline{u}}{\displaystyle\left(\sum_{n=0}^N \frac{\hat{\rho}^{2n}}{2n+1}\right)}$&$\mathcal{F}_\rho\overline{u}$&$\overline{p}$&$\mathcal{F}_{\rho}\left[\epsilon_{\mathrm{ref}}+
    \left(c_0\overline{T}^H-c_{-1}\right)
    \displaystyle{\sum_{n=0}^N \frac{\hat{T}^{2n}}{2n+1}}+
    c_{-1}\left(1+\overline{\log{T}}\right) + \xi\right]$&$\overline{\overline{(u,p)}}$\\
    \vspace{0.1cm}\\

        \citet{Gouasmi_CMAME_2020}&\ding{51} &\ding{51}&$\overline{\rho}^{\mathrm{log}}\overline{u}$&$\mathcal{F}_\rho\overline{u}$&$\dfrac{R\overline{\rho}}{\overline{1/T}}$&$\mathcal{F}_{\rho}\left[\epsilon_{\mathrm{ref}} +
    c_{-1}\left(1-\dfrac{\overline{1/T}}{\overline{1/T}^{\mathrm{log}}} + \overline{\log T}\right)+\dfrac{c_0}{\overline{1/T}^{\mathrm{log}}} + \xi
    \right]$&$\overline{u}\dfrac{R\overline{\rho}}{\overline{1/T}}$\\
    \vspace{0.1cm}\\
        \citet{Ranocha_JSC_2018}  & \ding{51} &\ding{55}&$\overline{\rho}^{\mathrm{log}}\overline{u}$&$\mathcal{F}_\rho\overline{u}$&$\overline{p}$&$\mathcal{F}_\rho\left[\overline{\left(\dfrac{1}{e}\right)}^{\mathrm{log}}\right]^{-1}$&$\overline{\overline{(u,p)}}$\\
        \vspace{0.1cm}\\
        KEEP \cite{Kuya_JCP_2018} & \ding{51} & {\ding{55}}&$\overline{\rho}\,\overline{u}$&$\mathcal{F}_\rho\overline{u}$&$\overline{p}$&$\mathcal{F}_\rho\overline{e}$&$\overline{\overline{(u,p)}}$\\
        \vspace{0.1cm}\\
        \bottomrule
    \end{tabular}}
    \caption{Summary of the compared numerical discretizations. \ding{51}: property verified for TP gases. {\ding{109}}: property verified for TP gases (asymptotically). \ding{55}: property not verified for TP gases. 
    For clarity purposes, we defined $\xi=\xi\left(\overline{\overline{T}}^m,\widetilde{\widetilde{T}}^m;c_m\right)=\sum_{m=-\ell}^{-2}\frac{c_m}{m+1}\widetilde{\widetilde{T}}^m +\sum_{m=1}^{r}\frac{c_m}{m+1}\overline{\overline{T}}^m$. We also reported the asymptotic approximation for the AEC-TP scheme, here reported as AEC$^{(N)}$-TP, since it has been  tested as well---although not reported in Section \ref{sec:results}---showing no relevant differences with respect to the logarithmic form, both with $N=3,5$. Further details on AEC formulation---for both CP and TP gases---can be found in~\cite{Aiello_JCP_2025,DeMichele_JCP_2023}.}
    \label{tab:schemes}
\end{sidewaystable}

All numerical simulations are carried out with sixth-order-accurate spatial discretization---for both convective and diffusive terms---and a third-order, low-storage explicit Runge--Kutta scheme~\cite{Wray1991} in the open-source, GPU-accelerated solver STREAmS 2.1~\cite{Salvadore2025}, where all the formulations in Table \ref{tab:schemes} have been implemented, besides the KEEP scheme which served as the built-in option. A fifth-order WENO scheme has been used to enhance stability during the initial transition to the turbulent state, until about $t=(R^\#T^\#_\mathrm{ref})^{1/2}\,t^\#/h^\#\approx25$, with the superscript “$\#$" indicating the dimensional quantities. Shock-capturing methods have been afterwards disabled and computations have been completed in a full-central framework with different numerical formulations of convective terms, reported in Table~\ref{tab:output} together with the main results of streamwise- and spanwise-averaged output quantities. The time integration is performed with an adaptive time step, with target CFL value of $ 0.6$.
Averaging procedure has been carried out after the transient stage had passed ($t\approx 400$) for each physical case. Then, about $N_s = 500$ samples have been collected, at constant spacing of $\Delta t_s = 1.5U_b/h$, to ensure both a sufficiently long observation window and an appropriate degree of decorrelation between snapshots. As a final note, we report that viscous terms have been discretized in their conservative form. Although non-conservative form is often employed to avoid odd-even decoupling effects~\cite{Bernardini2021}, we found, specifically for the numerical tests under scrutiny, more robust performances by discretizing divergence forms and therefore guaranteeing total-energy conservation.

\section{Numerical results}\label{sec:results}
This section presents turbulent channel flow simulations of carbon dioxide at high-supersonic and low-hypersonic regimes performed with the structure-preserving methods presented in Section \ref{sec:mathematical_formulation} to improve solution fidelity. Main physical features will be discussed, as well as the numerical differences arising from the use of different discretizations of the energy equation, showing that numerical methods which are specifically designed to enforce additional structural properties for TP gases give better overall performances. In particular, we selected the numerical schemes to compare according to specific criteria: KEEP formulation is one of the most widely used schemes and serves as the baseline choice in the STREAmS code, representing a highly robust KEP scheme for CP gases, although not strictly EC for both CP and TP gases.
On the other hand, the scheme proposed by \citet{Gouasmi_CMAME_2020} is EC for thermally perfect gases. It is also KEP according to our definition, despite adopting a discretization of the pressure terms that has been shown to spoil the kinetic-energy balance in homogeneous isotropic turbulence cases~\cite{Gassner_JCP_2016,Aiello_ArXiv_2025} (similar to the scheme proposed by \citet{Chandrashekar_CCP_2013}, to which it reduces in the CP gas case).
Finally, for the hypersonic case, an additional comparison was made using the scheme of Ranocha~\cite{Ranocha_JSC_2018}, which is exactly EC for the CP gas case, intended to offer intermediate performances between KEEP and schemes specifically designed for thermally perfect gases. The latter statement will be assessed by the reported results, showing both the role of the EC property and the consistency with the thermodynamic model.
\begin{table}
\centering
\normalsize
\renewcommand{\arraystretch}{1.2}
\begin{tabular}{c c c c c c c c c c c}
\toprule
     & $M_b$  &\Rey$_{b}$& $\Rey_\tau$ & $\Delta x^+$ & $\Delta y^+_w$ & $\Delta z^+$ & $N_x$ & $N_y$ & $N_z$  \\ \hline

 \textbf{SCM3} & 3.0 &$\sim$5860& $\sim470$ & 9.18 & 0.28 & 4.35 & 512 & 256 & 324  \\
\textbf{SCM4} & 4.0 &$\sim $6850& $\sim630$ & 8.21 & 0.31 & 3.93 & 768 & 312 & 480 \\
\textbf{HCM5} & 5.0 &$\sim $7460& $\sim800$ & 9.68 & 0.32 & 4.62 & 816 & 356 & 512 \\
\bottomrule
\addlinespace
\end{tabular}
\caption{Numerical parameters for the DNS of the supersonic/hypersonic channel flow. Main physical parameters (bulk Mach number $M_b$, bulk Reynolds number $Re_b$, friction Reynolds number $Re_\tau$), inner-scaled grid spacings $\Delta x^+, \Delta y^+_w, \Delta z^+$ and number of grid points $N_x,N_y,N_z$ for the supersonic channel flow at $M_b=3,\,4$ (SCM3,SCM4) and the hypersonic channel flow at $M_b=5$ (HCM5).}
\label{tab:inputpar}
\end{table}

Table~\ref{tab:inputpar} summarizes the selected physical cases along with the main physical and grid parameters. All simulations were carried out in a computational box of size $\Omega=[0,L_x]\times[0,L_y]\times[0,L_z]=[0,10h]\times[0,2h]\times[0,3h]$, inspired by previous numerical work on compressible channel flows~\cite{Trettel2016,hasan2025}. Based on previous grid sensitivity analysis, the selected domain and meshes appear adequate for the scope of the analysis presented in the next sections, which basically revolves around Favre-averaged velocity and temperature distributions and second-order turbulent statistics.

\subsection{Mean flow statistics}

\begin{figure}[t]
  \centering
  \subfloat[]{\includegraphics[width=.33\textwidth]{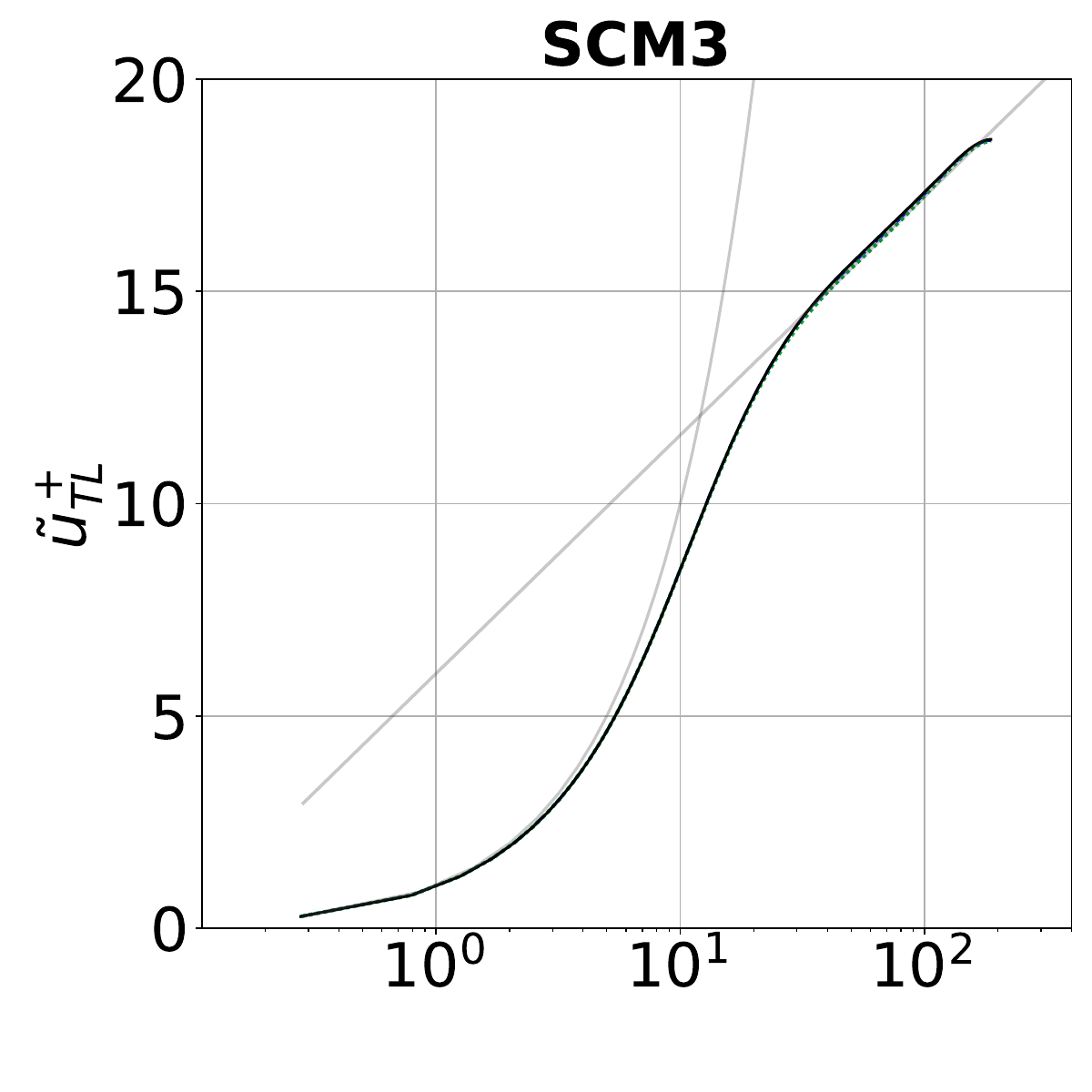}}
  \subfloat[]{\includegraphics[width=.33\textwidth]{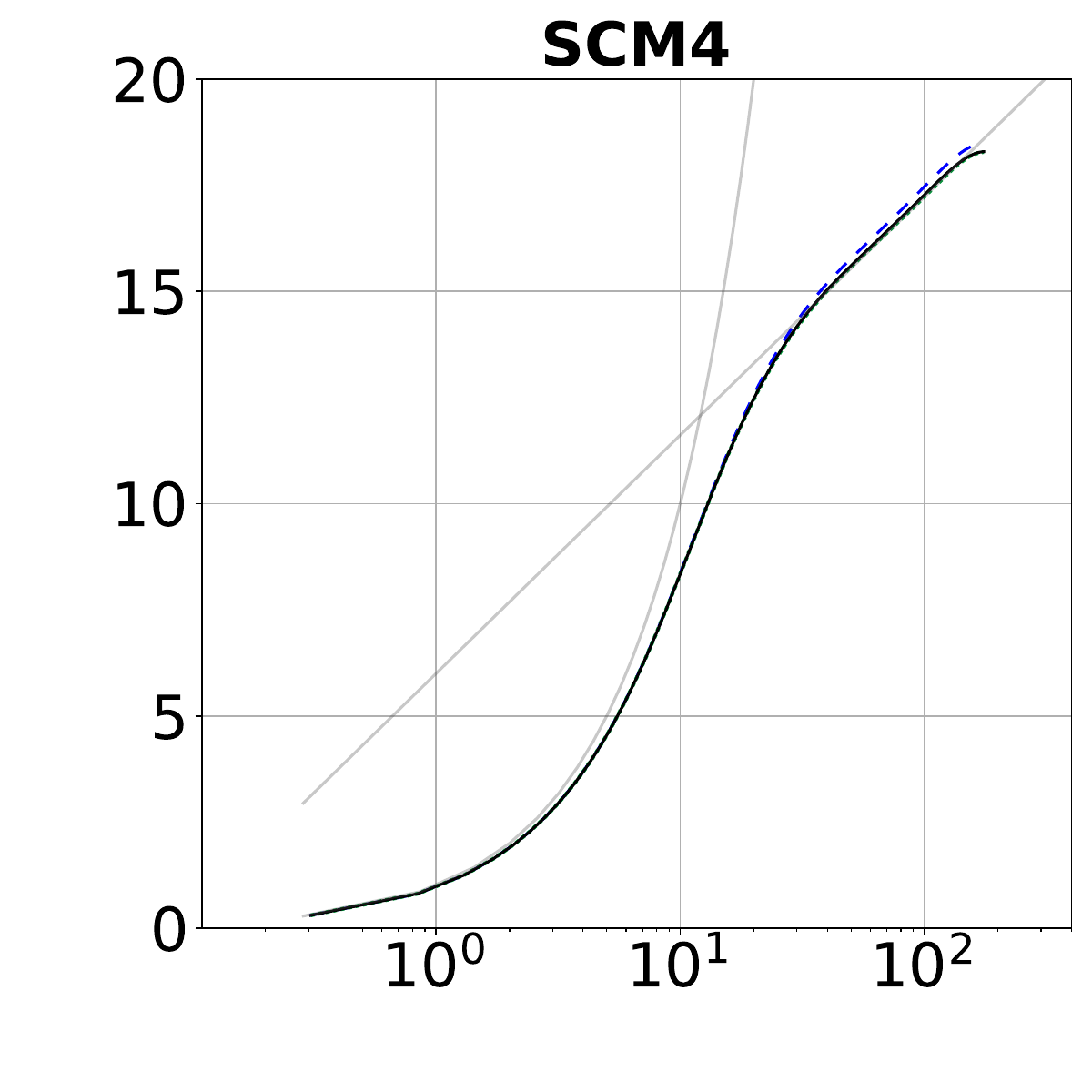}}
  \subfloat[]{\includegraphics[width=.33\textwidth]{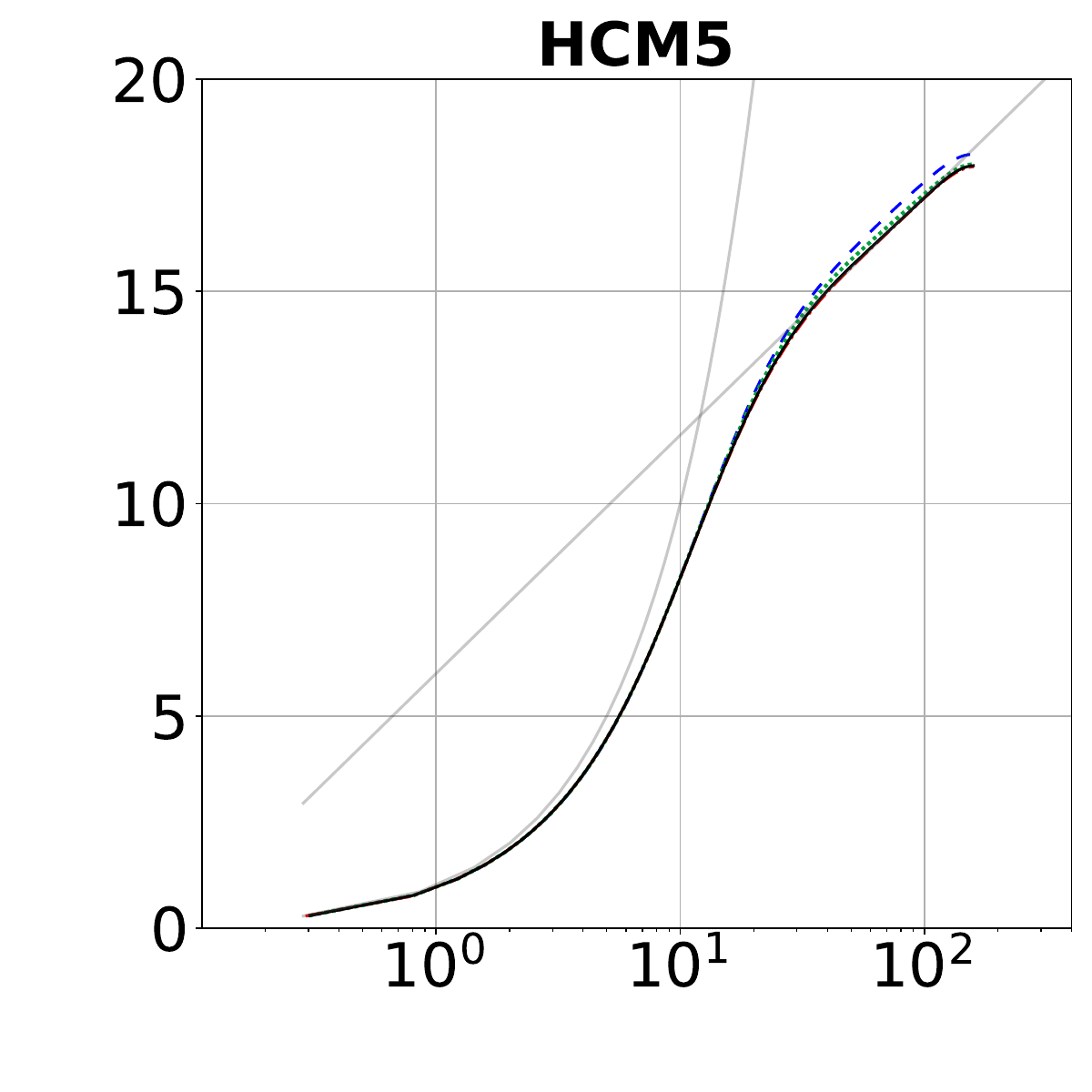}}\\
  \subfloat[]{\includegraphics[width=.33\textwidth]{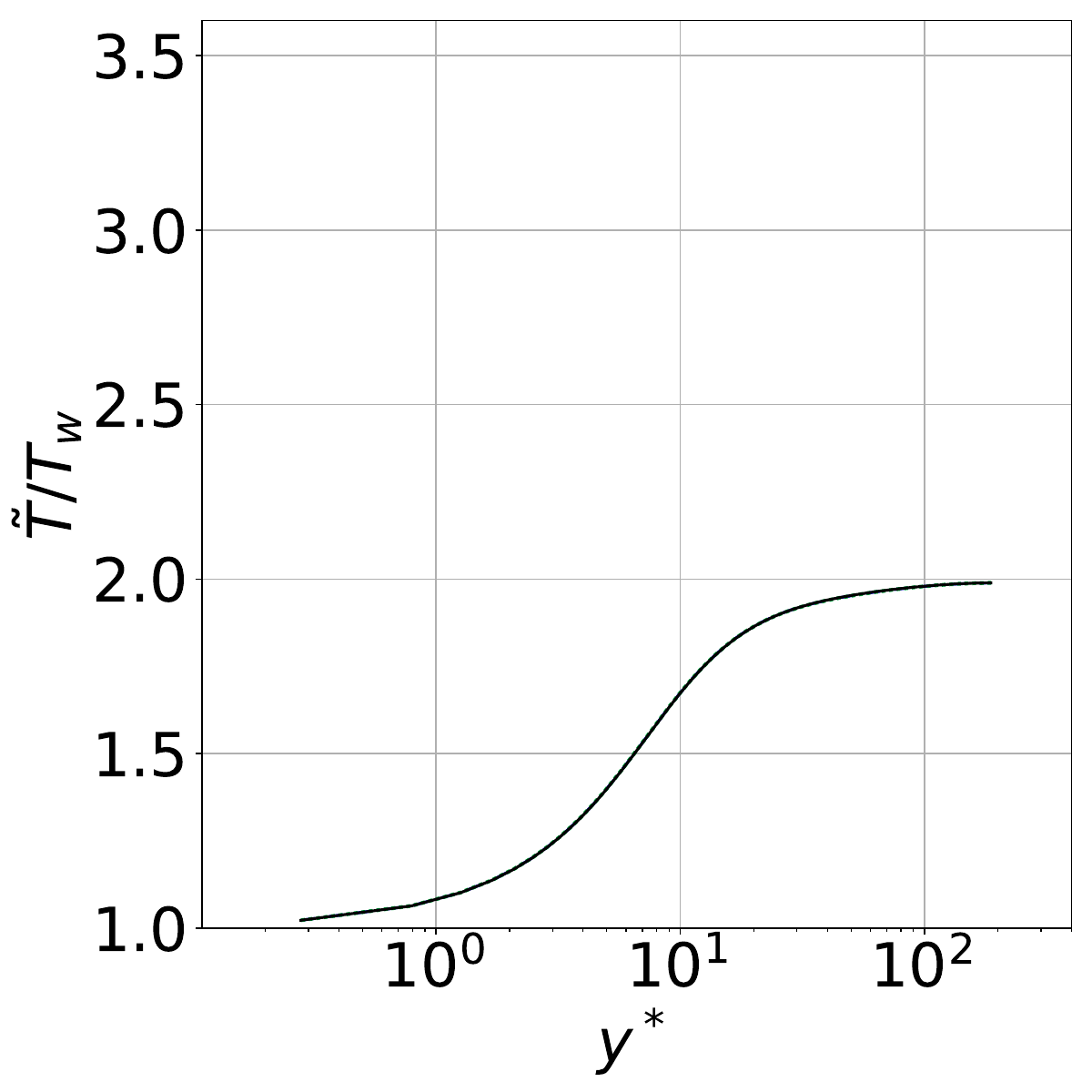}}
  \subfloat[]{\includegraphics[width=.33\textwidth]{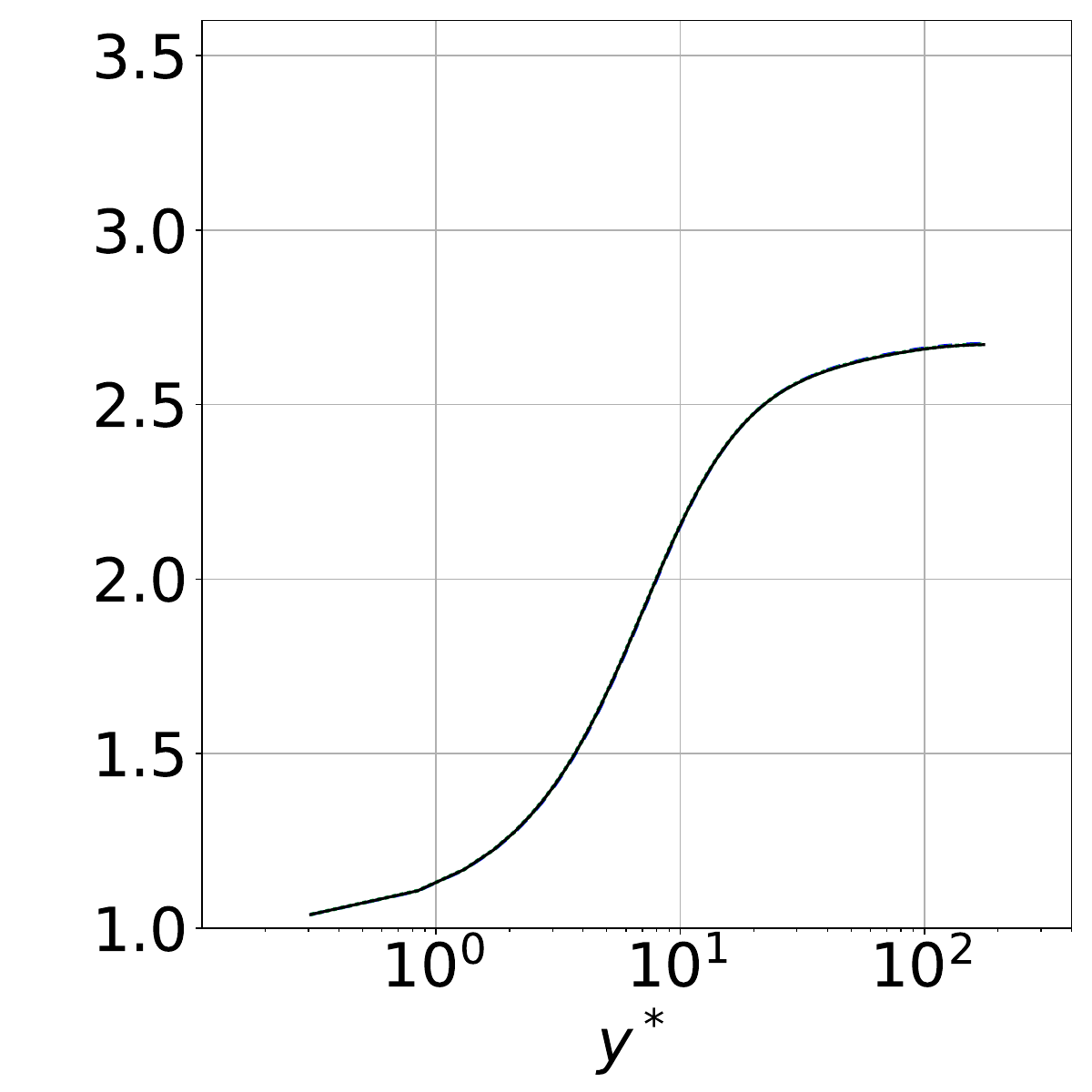}}
  \subfloat[]{\includegraphics[width=.33\textwidth]{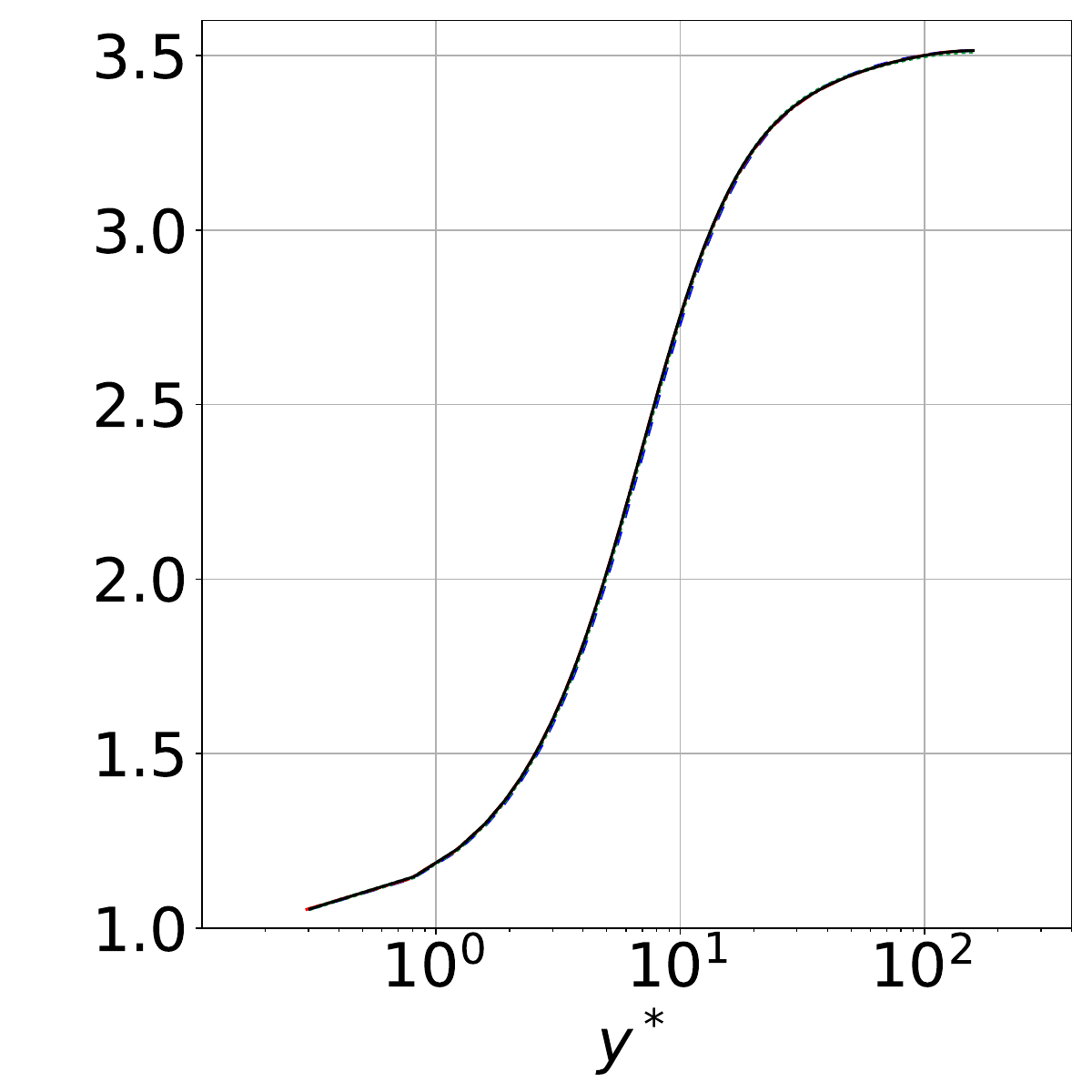}}
  \caption{Wall-normal distributions of inner-scaled, Trettel--Larsson transformed, Favre-averaged streamwise velocity component $\tilde{u}^+_{TL}$ (panels (a)--(c)) and temperature $\tilde{T}$ (panels (d)--(f)) profiles, in semi-local scaling $y^*$. Solid black lines: EC-TP; dashed blue lines: KEEP; dotted green lines: Gouasmi et al.; dash-dotted red lines: Ranocha. In panels (a)--(c), solid gray curves represent standard law-of-the-wall trends in the viscous sublayer $u^+=y^+$ and the log-law $u^+=6.0+1/0.41\log{ y^+}$.}
   \label{fig:velprofiles}
\end{figure}

Fig.~\ref{fig:velprofiles} shows wall-normal distributions of Favre-averaged velocity and temperature profiles for all the numerical tests reported in Table~\ref{tab:schemes}, in terms of the semi-local wall-normal distance $y^*$. Velocity data are reported using the Trettel--Larsson (TL) transformation~\cite{Trettel2016}

$$ u^+_{TL} = \int_0^{u^+} \left(\dfrac{\langle\rho\rangle}{\rho_w}\right)^{1/2}\left(1+\dfrac{1}{2\langle\rho\rangle}\dfrac{\mathrm{d}\langle \rho\rangle}{\mathrm{d}y}y-\dfrac{1}{\langle\mu\rangle}\dfrac{\mathrm{d}\langle \mu\rangle}{\mathrm{d}y}y \right) du^+$$
which, for CP gases, is known to collapse supersonic channel velocity data-sets onto their incompressible counterpart. Regardless of the numerical method used and the Mach number, inspection of Fig.~\ref{fig:velprofiles} (a)--(c) reveals that all TL-transformed profiles land on the incompressible wall law in the viscous sublayer, say $\tilde{u}_{TL}^+ = y^*$. In addition, a narrow logarithmic region appears for $y^* \geq 40 $, which well conforms with the classical log-law trend $u ^ + = 1/\kappa \log (y^*) + B$, with $\kappa = 0.41$, and $B = 6.0$ being calculated fitting the EC results. The value of $B$ found for the present data-set is related to the small transformed Reynolds number $Re_\tau^*$ of the simulations (ranging from 160 and 190 in the present simulations). The larger log-law intercept of $\tilde{u}^+_{TL}$ is in accordance with the behavior of low $Re_\tau$ incompressible flows and was already observed in supersonic channel flow simulations of CP gases~\cite{Trettel2016,Modesti2016}.

While the above considerations appear to be well suited for all Mach numbers when the EC discretization is used, a careful look at
Fig.~\ref{fig:velprofiles} (a)--(c) reveals that the numerical discretization can have some influence on the numerical results. Specifically, as the Mach number increases, differences in the logarithmic region arise, with the KEEP scheme appearing to overpredict the value of the TL-transformed velocity for high $y^*$. This fact well conforms with the theoretical considerations made in Section \ref{sec:mathematical_formulation}, which allow to state that the KEEP scheme is the most inaccurate one. The EC formulation of Gouasmi et al.~exhibits a similar behavior with the Mach number, as it slowly departs from the EC-TP for the HCM5 case, although showing less pronounced discrepancy.
A quantitative evaluation of the differences among the various formulations is reported in the columns $\mathrm{r}(\tilde{u}^+)$ and  $\mathrm{r}(\tilde{u}^+_{TL})$ in Table~\ref{tab:output}, where the maximum normalized relative differences are reported for the non-transformed and TL-transformed velocity profiles.

On the other hand, Fig.~\ref{fig:velprofiles}(d)--(f) report the mean temperature profiles. As expected, the ratio between the centerline and wall temperatures increases with Mach number, as a consequence of the balance between heat fluxes and viscous dissipation. For the HCM5 case, the centerline temperature reaches approximately 1000 K, thereby amplifying the differences between the CP and the TP model results.

\subsection{Turbulent fluctuations}

\begin{figure}
    \centering
    \subfloat[]{\includegraphics[width=.33\textwidth]{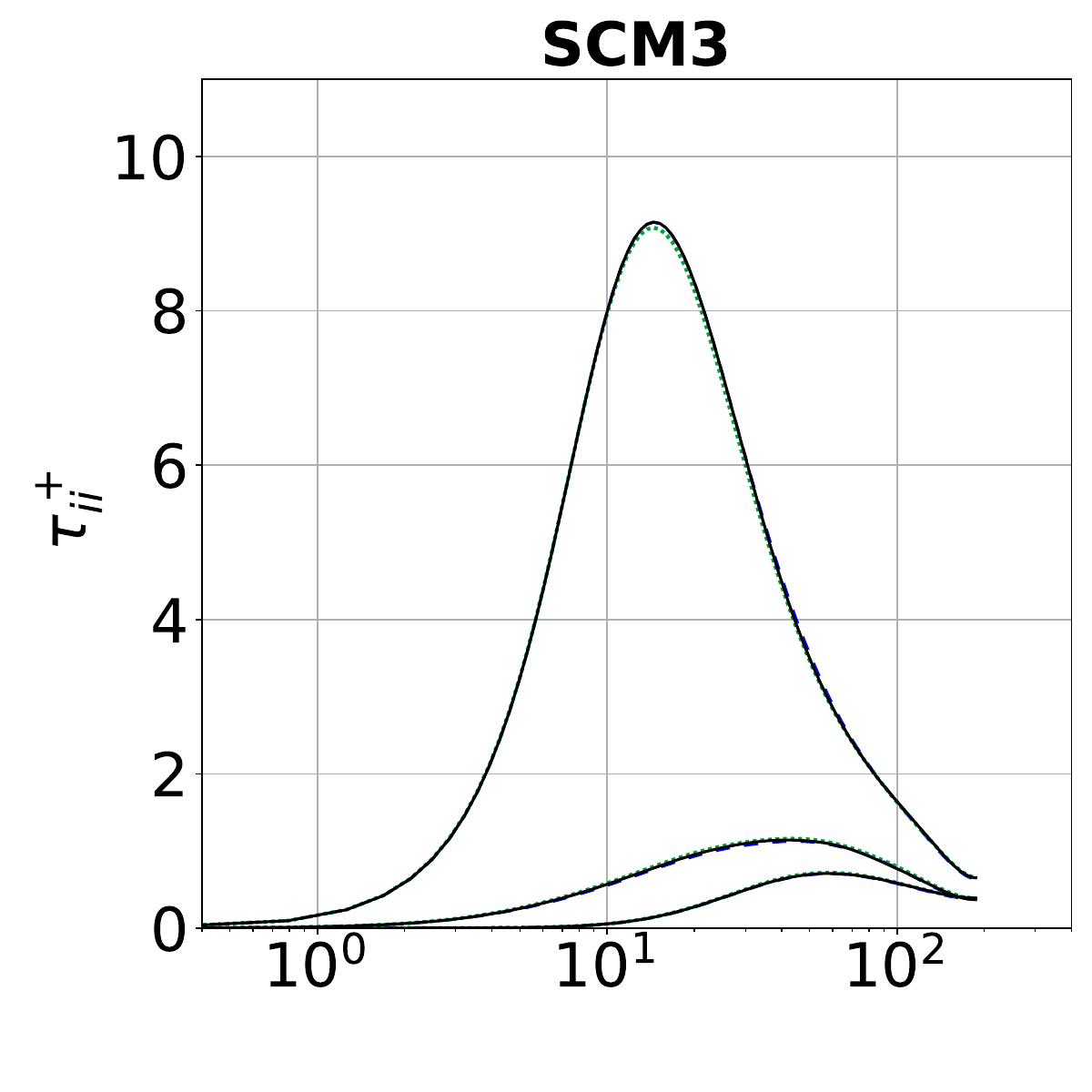}}
  \subfloat[]{\includegraphics[width=.33\textwidth]{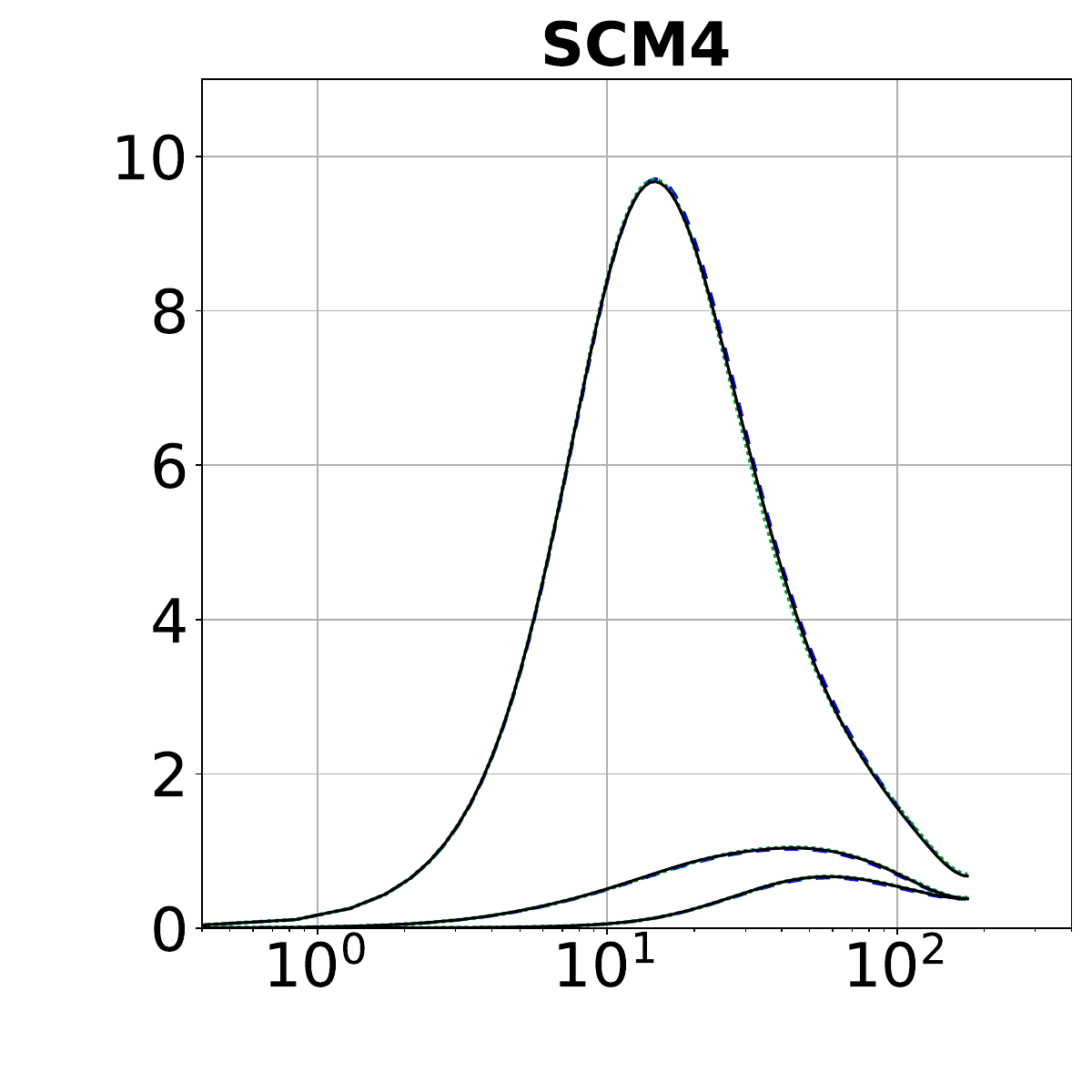}}
  \subfloat[]{\includegraphics[width=.33\textwidth]{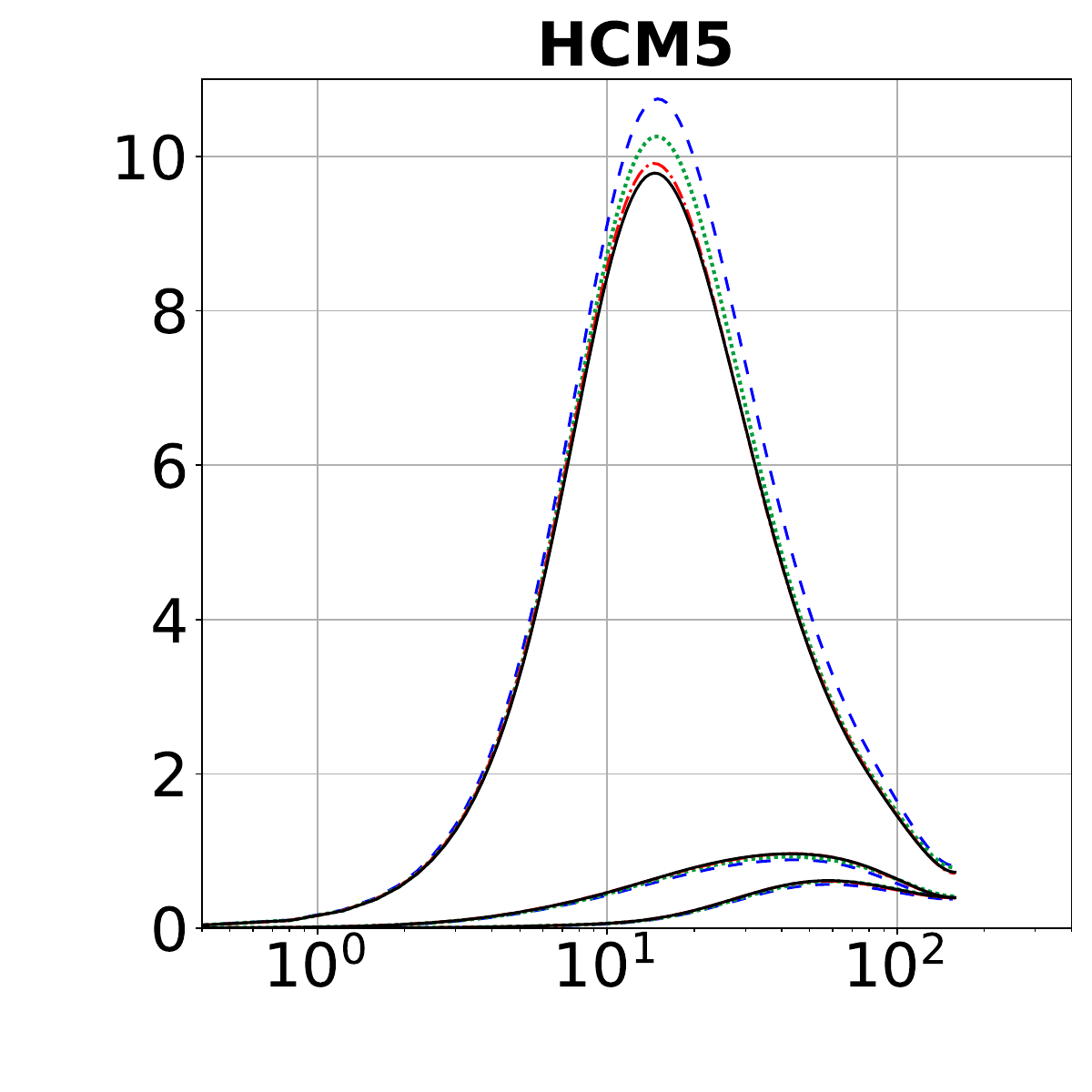}}\\

   \subfloat[]{\includegraphics[width=.33\textwidth]{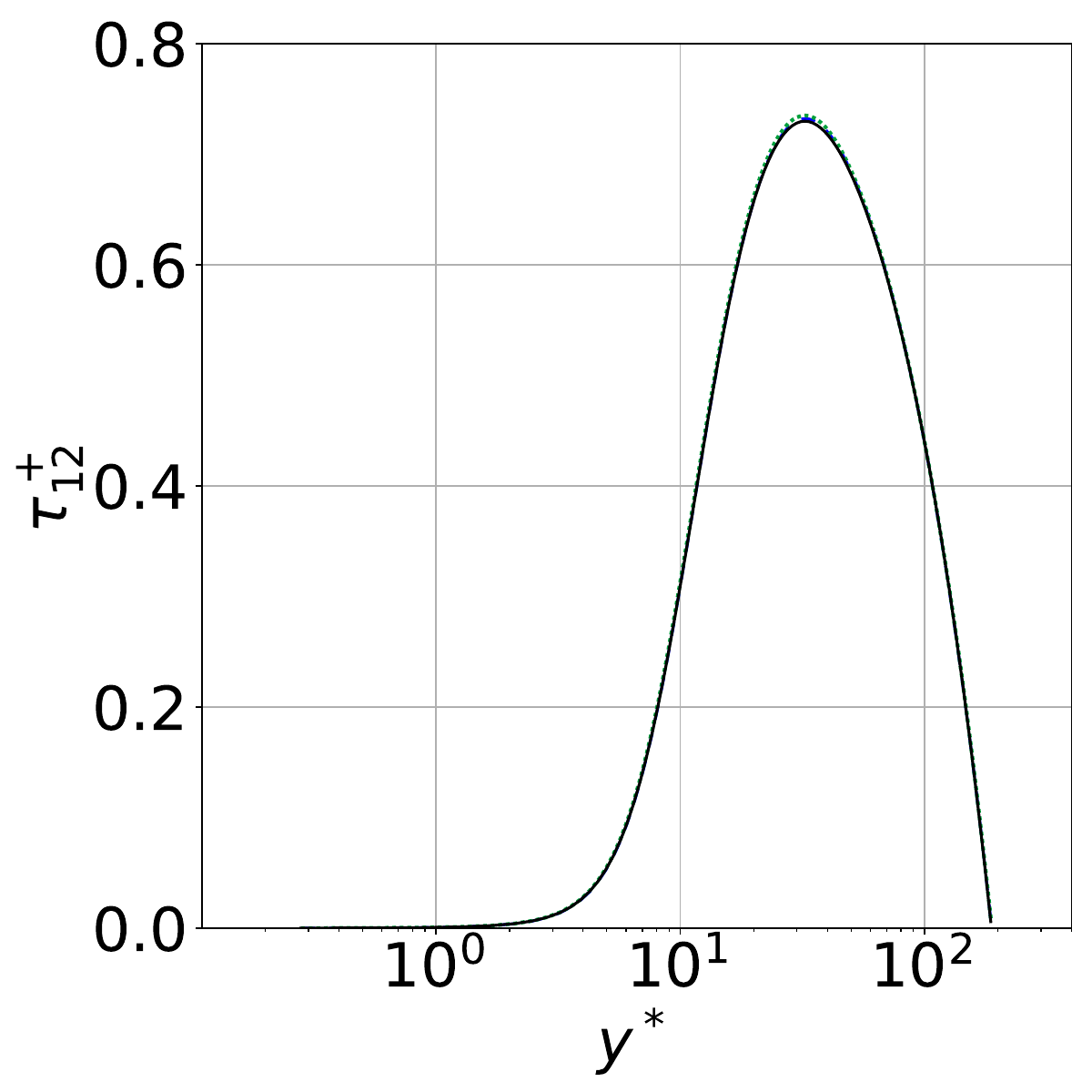}}
  \subfloat[]{\includegraphics[width=.33\textwidth]{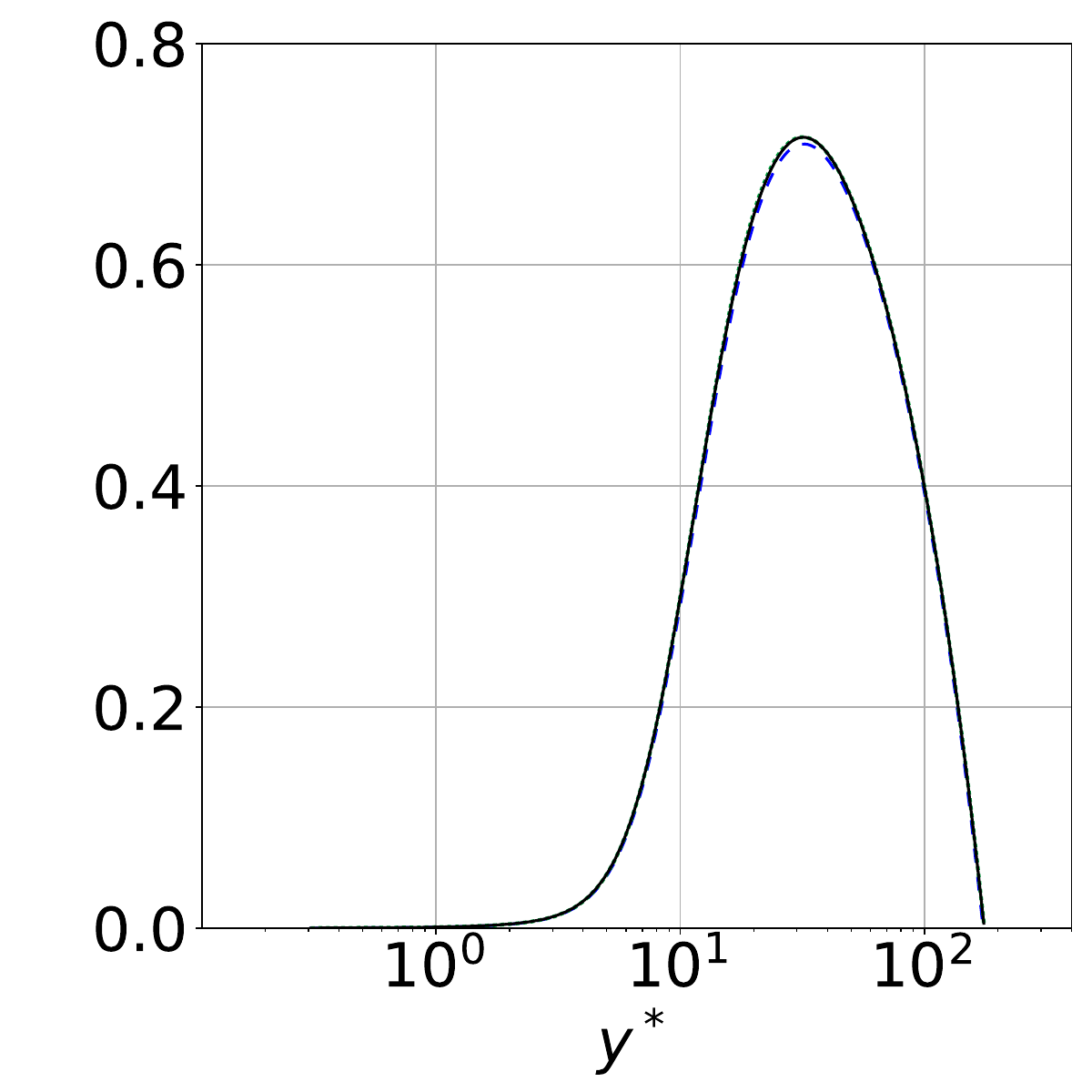}}
  \subfloat[]{\includegraphics[width=.33\textwidth]{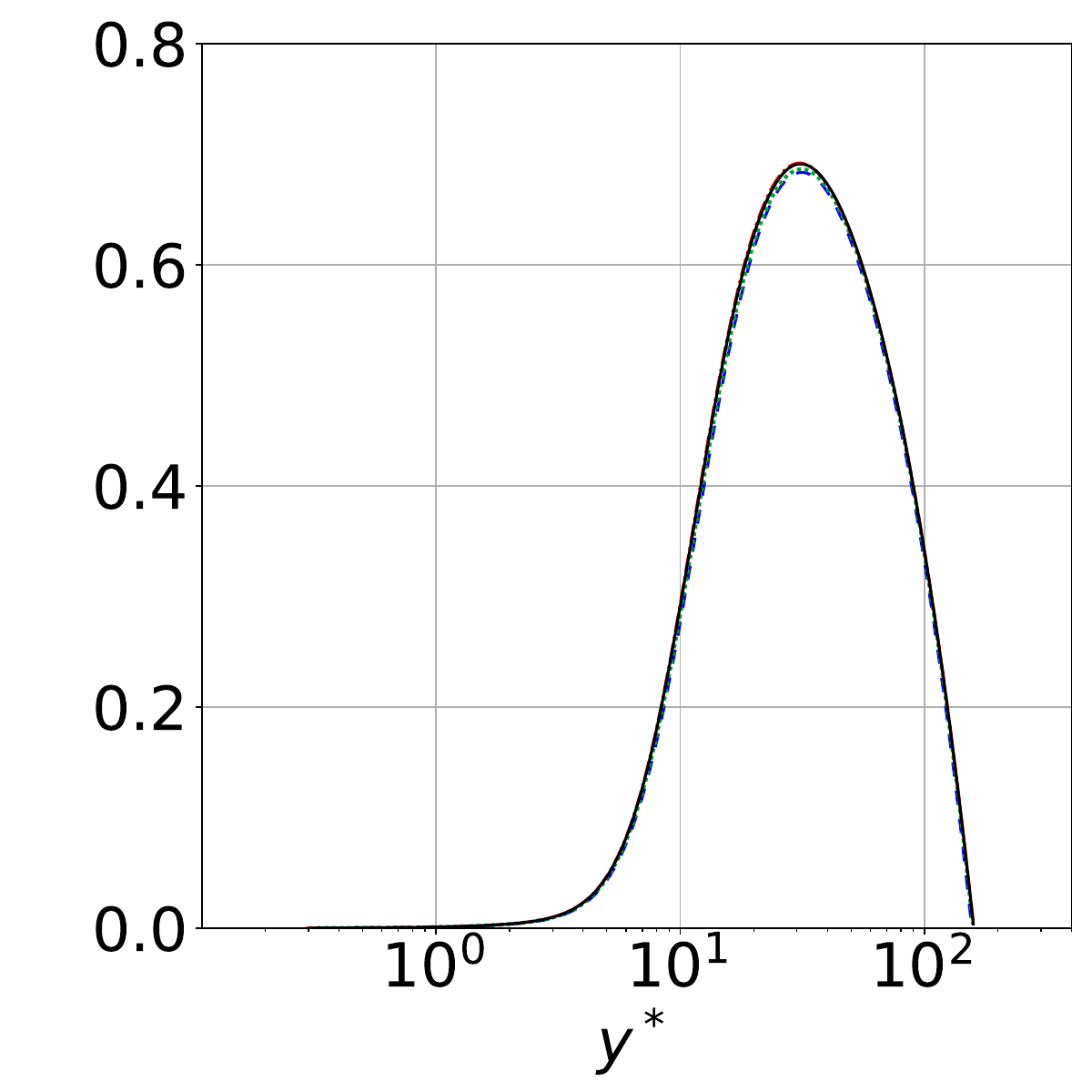}}
    \caption{Wall-normal distributions of the density-scaled  turbulent stresses $\tau_{ij}^+ =  \langle\rho\rangle \widetilde{u''_{i} u''_{j}}/\tau_w$ in semi-local units $y^*$. Colors are set as in Fig.~\ref{fig:velprofiles}. Panels (a)--(c) display the orthogonal components of the turbulent stresses at various Mach numbers, whereas panels (d)--(f) display turbulent shear stress $\tau_{12}^+$.}
    \label{fig:reystress}
\end{figure}

The results of the comparative analysis made in the previous section on the Favre-averaged data-sets appears to be well confirmed by the trends of the Reynolds stresses (Fig.~\ref{fig:reystress}), which are reported in semi-local scaling as
\begin{equation*}
    \widetilde{u''_iu''_j}^* = \frac{\langle\rho\rangle \widetilde{u''_iu''_j}}{ \tau_w}= \tau^+_{ij}.
\end{equation*}
Indeed, The KEEP formulation shows abnormally high maxima for the streamwise Reynolds stress $\tau_{11}^+$, especially in the hypersonic case, again suggesting an overall inaccuracy arising from the specific discretization of the convective contributions. Comparison of results in Fig.~\ref{fig:reystress} (a)--(c) is particularly interesting when supported by the discretization features cataloged in Table~\ref{tab:schemes}: while the EC-TP scheme features an entropy-conservative convective part and a pressure term of the momentum equation discretized as $\mathcal{P}_{\rho u}=\overline{p}$, the formulation by Gouasmi et al.~retains the same convective discretization but employs a different treatment of the pressure term. The KEEP discretization, on the other hand, uses the same pressure discretization as the EC-TP scheme, while its convective part is not EC for TP gases. We can therefore isolate these numerical effects and infer that both the absence of an entropy-conservative convective term and the use of a pressure term as ${\mathcal{P}}_{\rho u}=R\overline{\rho}/\overline{(1/T)}$ lead to an inaccurate reconstruction of the correct physical trends. It is worth noting that pressure term discretizations as that used by Gouasmi et al.~exhibited limitations in reconstructing the energy-exchange mechanisms even in the inviscid case of homogeneous isotropic turbulence~\cite{Aiello_ArXiv_2025}. Those aspects both affect the performance of the numerical methods, and their optimal combination appears to be that provided by the EC-TP scheme, for which the velocity profiles and Reynolds stresses show improved behaviors. In contrast, no relevant differences have been found regarding the scaled turbulent shear stress $\tau_{12}^+$, shown in Fig.~\ref{fig:reystress} (d)--(f), for all the considered schemes and Mach numbers.

As differences among the numerical methods employed in this work lie in the choice of suitable discretizations of the energy equation, and since the enforcement of the EC property is dependent on the specific gas model, it appears natural to examine the impact of the numerical treatment on the thermodynamic quantities. In fact, the imposed equilibrium between the viscous stresses and the heat transfer in the present turbulent channel flow configuration leads to higher wall heat transfer when the Mach number increases. This fact, as previously pointed out, results in higher centerline-to-wall temperature ratios $ T_c /T_w$ for increasing Mach number, which are responsible for larger temperature changes across the wall-normal direction (as shown in Fig.~\ref{fig:velprofiles}(d)--(f)), and hence stronger compressibility effects. It is worthwhile remarking that this feature also exacerbates non-linear internal energy trends in the present configuration, hence we expect noticeable differences for the thermodynamic flow properties.

\begin{figure}
    \centering

     \subfloat[]{\includegraphics[width=.33\textwidth]{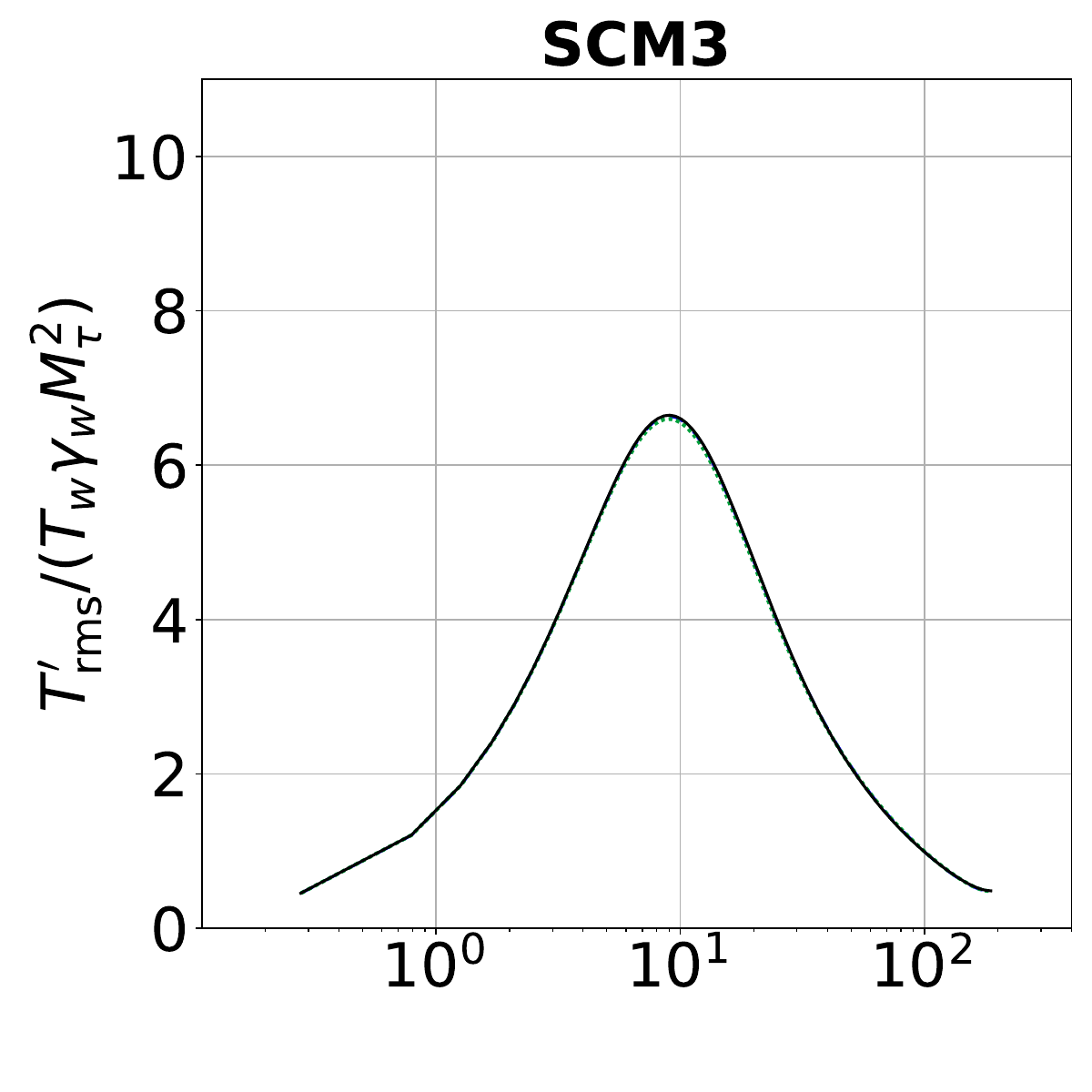}}
  \subfloat[]{\includegraphics[width=.33\textwidth]{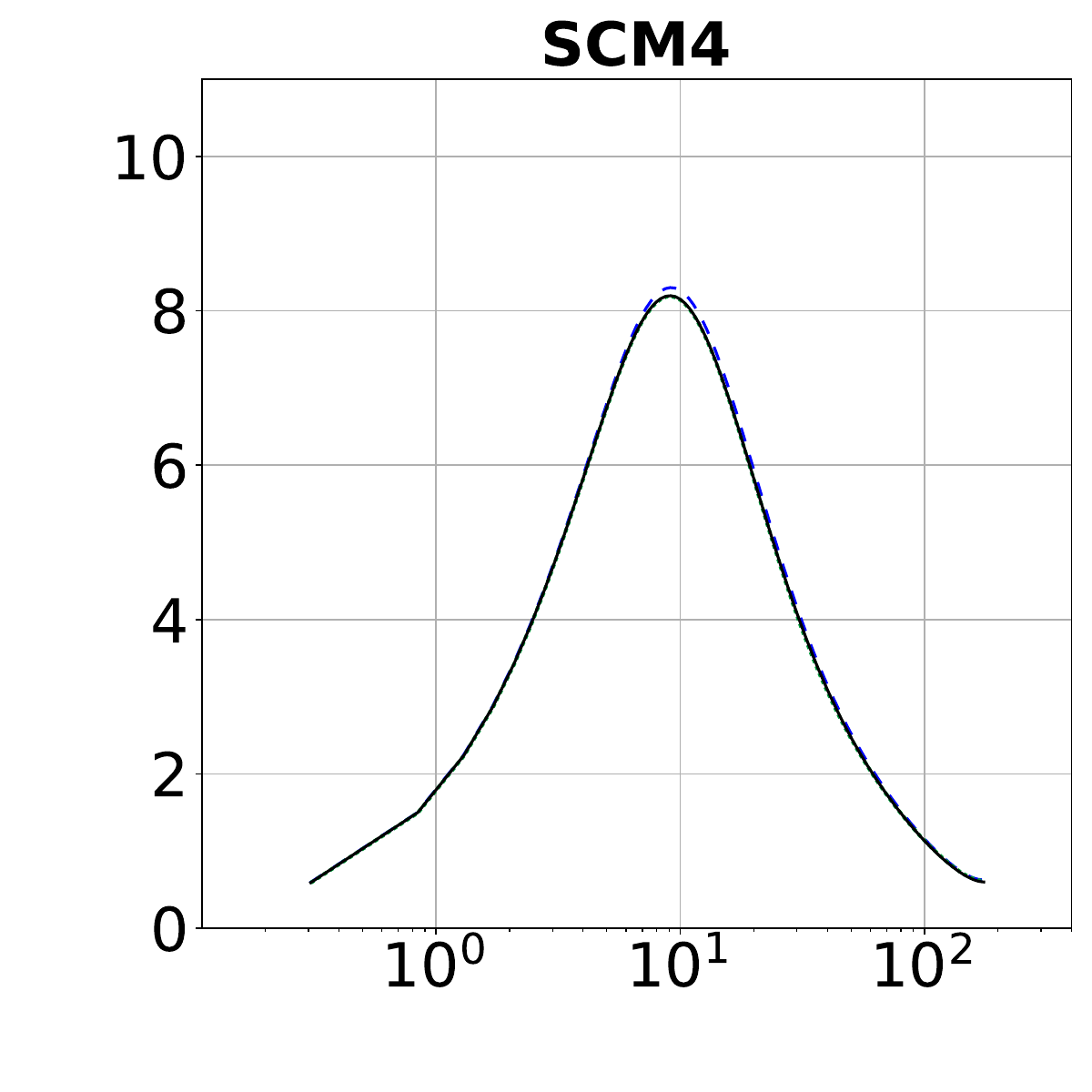}}
  \subfloat[]{\includegraphics[width=.33\textwidth]{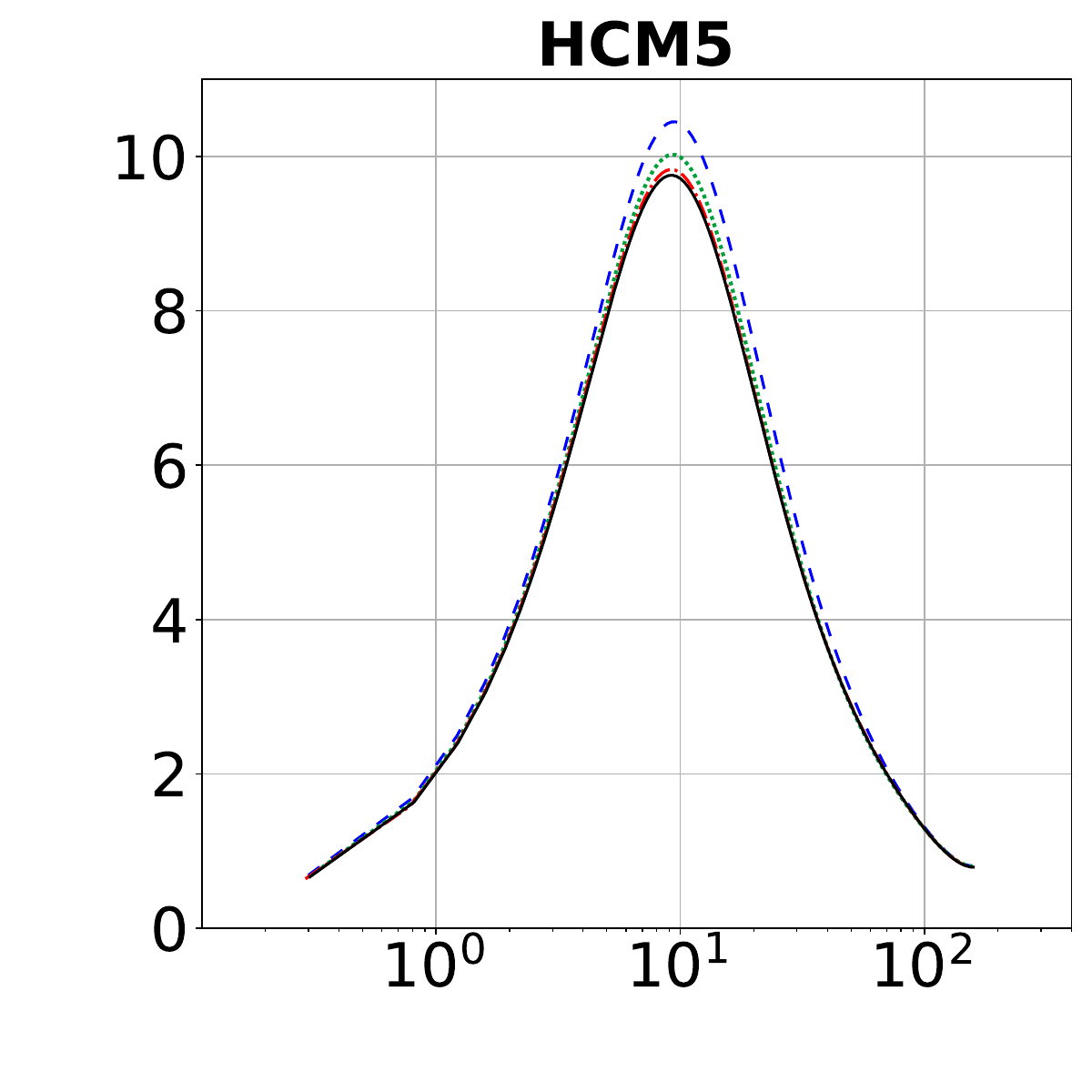}}\\
     \subfloat[]{\includegraphics[width=.33\textwidth]{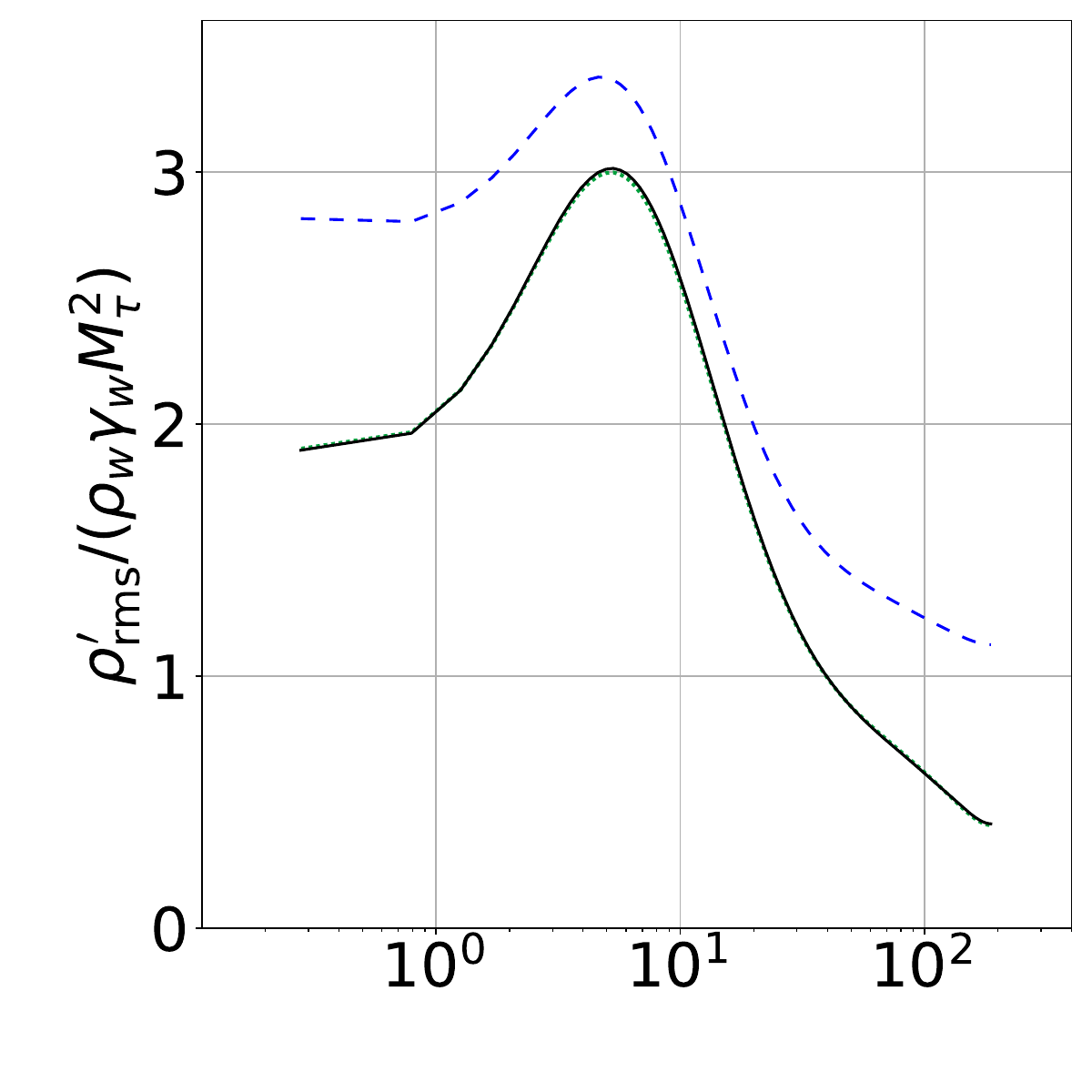}}
  \subfloat[]{\includegraphics[width=.33\textwidth]{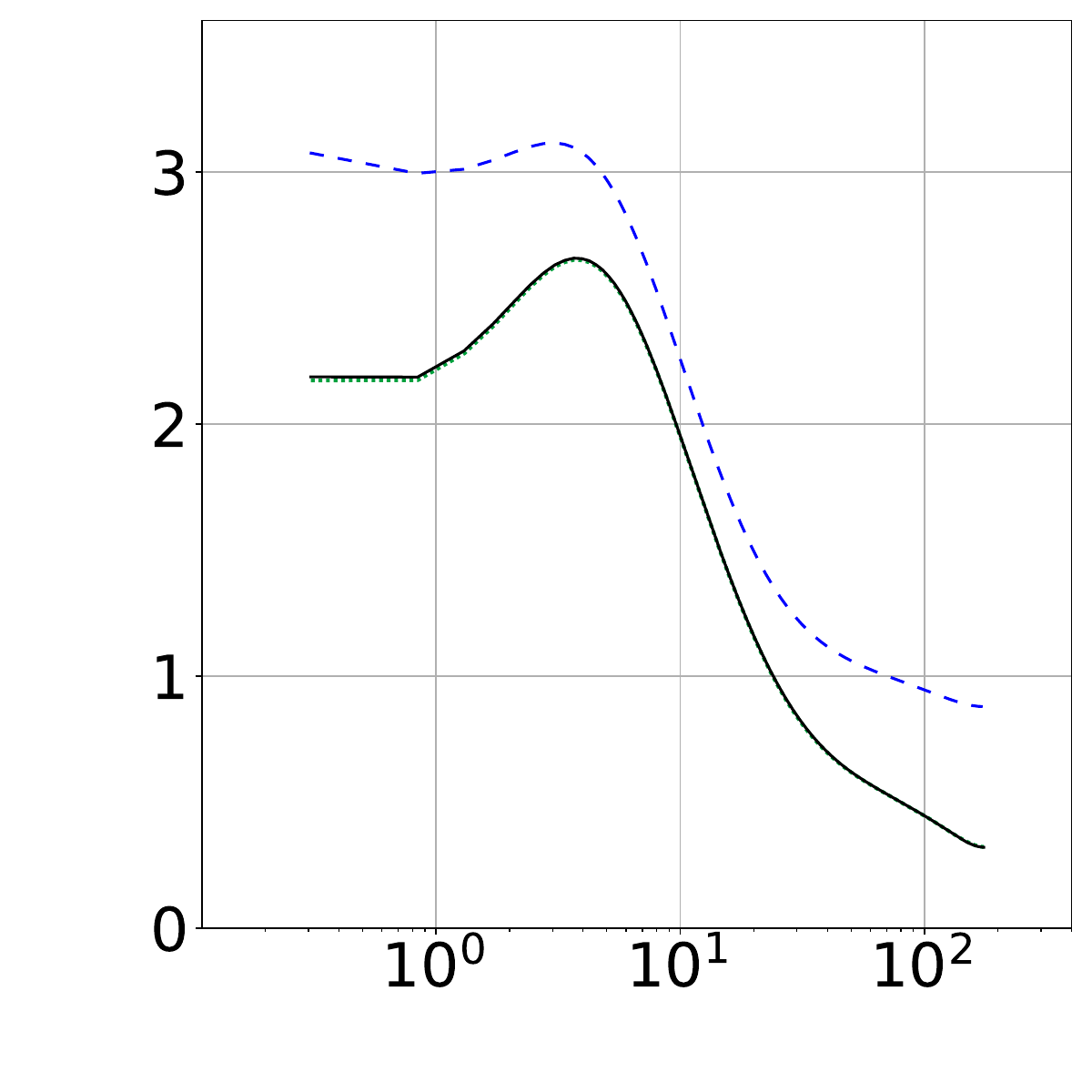}}
  \subfloat[]{\includegraphics[width=.33\textwidth]{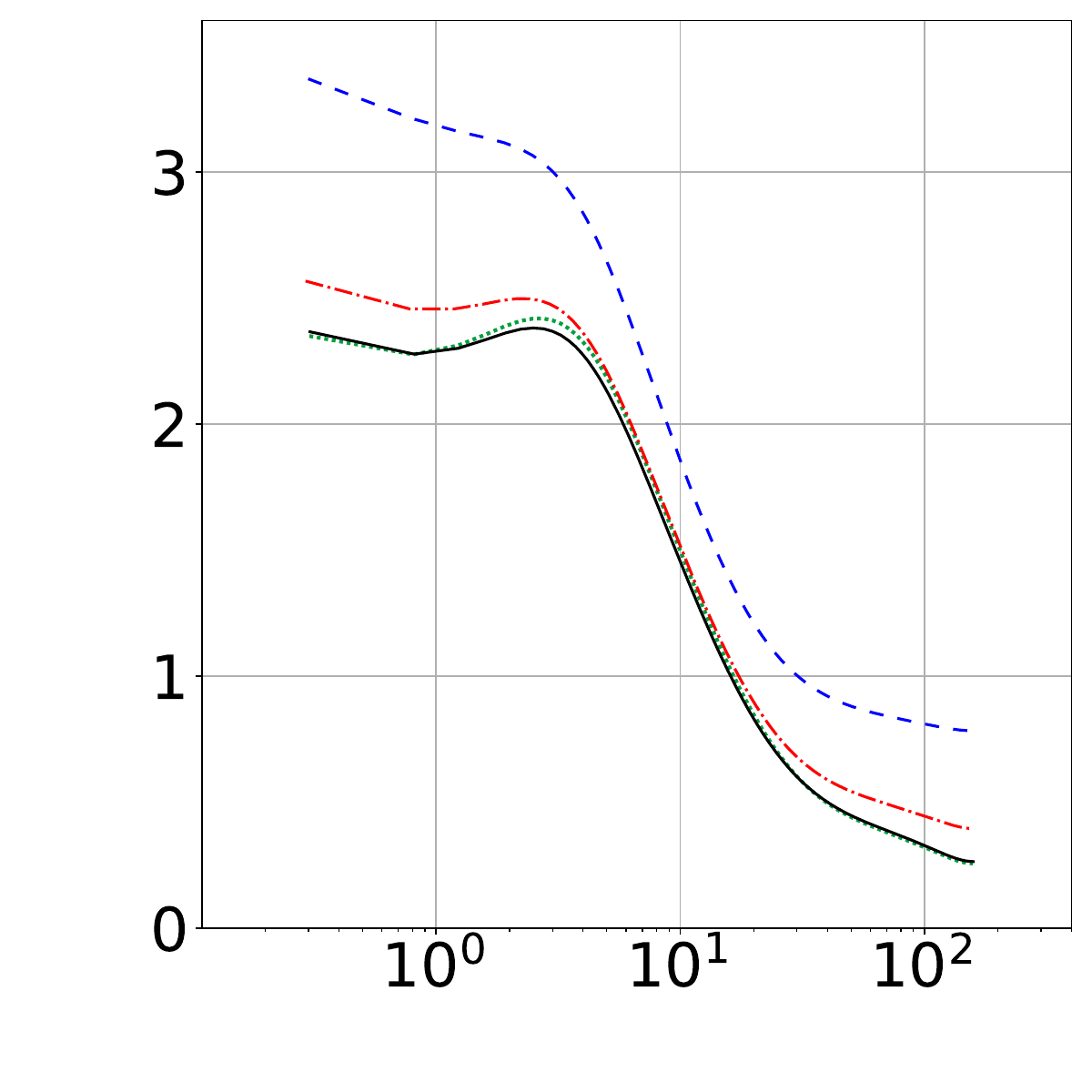}}\\
    \subfloat[]{\includegraphics[width=.33\textwidth]{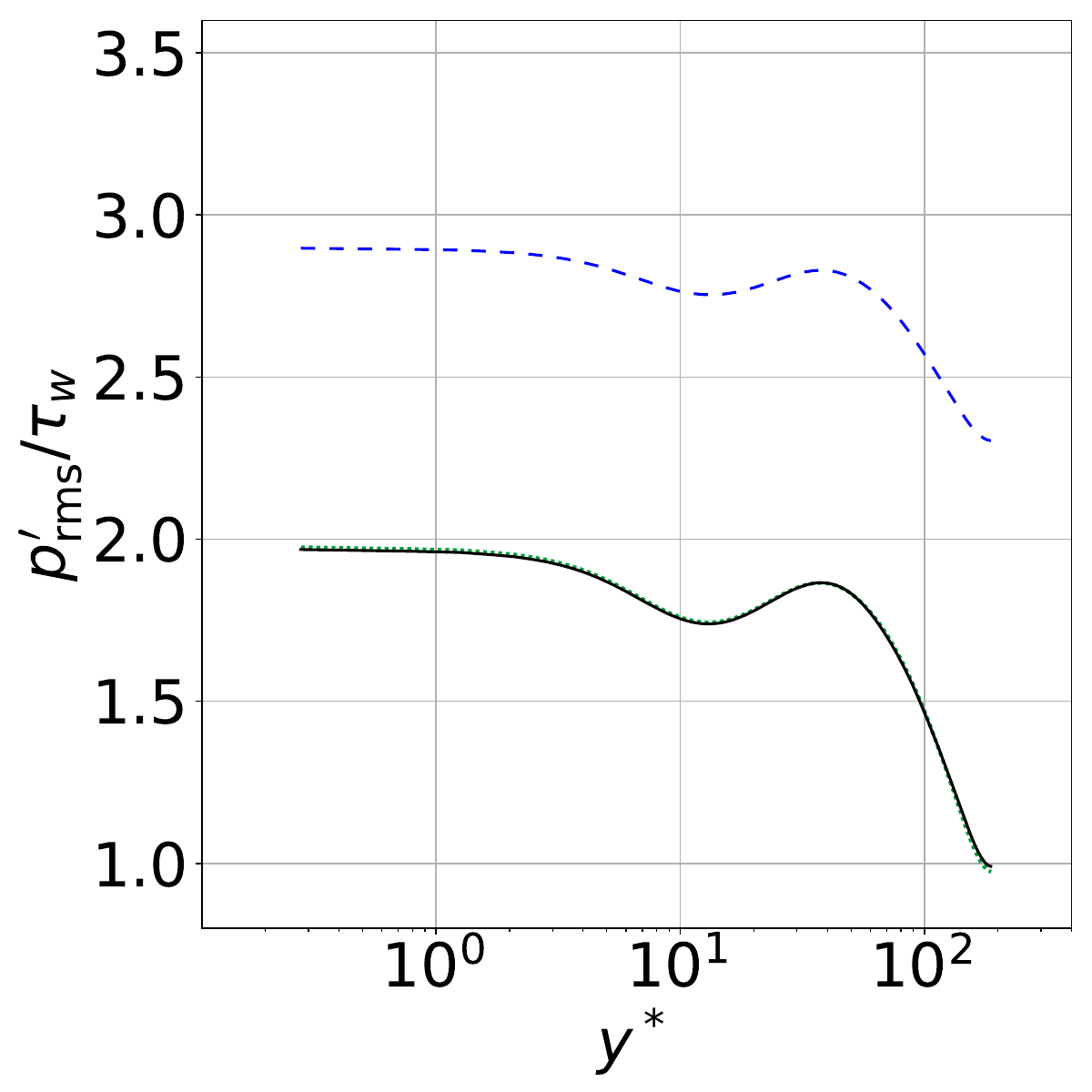}}
  \subfloat[]{\includegraphics[width=.33\textwidth]{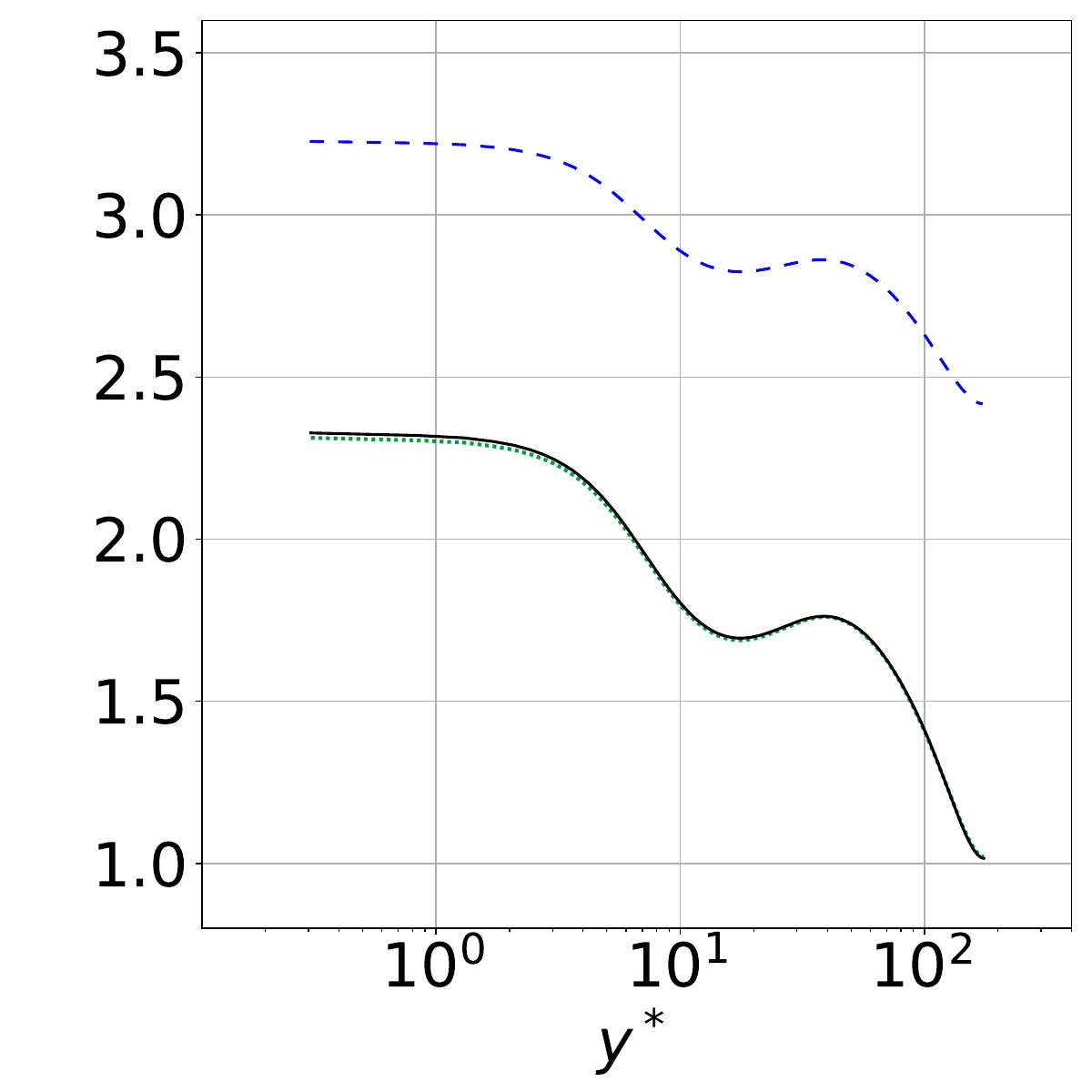}}
  \subfloat[]{\includegraphics[width=.33\textwidth]{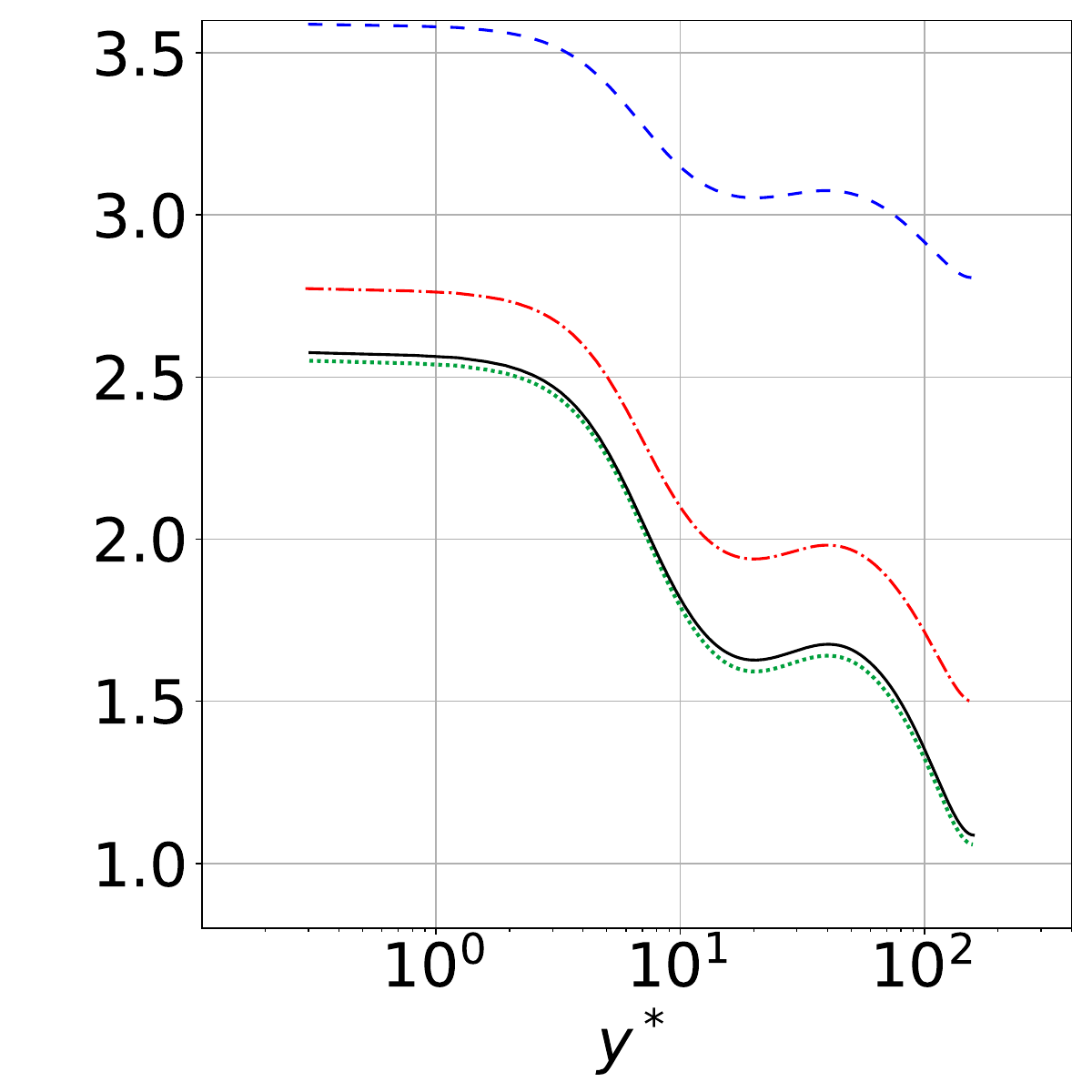}}\\
    \caption{Wall-normal distributions of normalized thermodynamic variable root-mean-squares in semi-local units $y^*$. Colors are set as in Fig.~\ref{fig:velprofiles}. Panels (a)--(c) display normalized temperature fluctuations, panels (d)--(f) report normalized density fluctuations, panels (g)--(i) depict inner-scaled pressure fluctuation  for various Mach numbers and numerical schemes.}
    \label{fig:thermo}
\end{figure}
Fig.~\ref{fig:thermo} reports wall-normal distributions of the main thermodynamic fluctuations in terms of root-mean-squares as $\phi^{\prime}_{\mathrm{rms}}=[N_s^{-1}\sum_{k=1}^{N_s}(\phi_k-\langle\phi\rangle)^2]^{1/2}$, and both numerical and statistical trends are clearly recognized.
Non-EC formulations appear to exhibit systematic overprediction in thermodynamic fluctuations, particularly in pressure and density.
Specifically, after a sufficiently long time following the attainment of a statistically stationary state, these schemes produce fluctuations that begin to drift, eventually leading to simulation blow-up for coarser grids (not reported here). To ensure that this issue did not affect all tested schemes, ruling out upstream causes, all simulations were run for an additional 50\% of the time reported at the beginning of this section. Nonetheless, both EC formulations maintained their statistically stationary state throughout.

Fig.~\ref{fig:thermo} also provides a possible interpretation of the previously discussed dynamical quantities: as Mach number increases, and particularly in the hypersonic regime, compressibility effects strengthen the thermodynamic–dynamic coupling for the Navier--Stokes system. In turn, numerical treatment of the energy equation gains increasing relevance and, at the discrete level, this results in a more pronounced influence on the velocity field---both averaged and fluctuating---which is indeed reflected in Fig.~\ref{fig:velprofiles}(a)--(c) and Fig.~\ref{fig:reystress} as a marked change in trend between the supersonic and hypersonic cases.

Furthermore, as thermodynamic fields become more relevant to the overall solution, so does the TP gas model and consequently the role of the EC property itself, boosting stability and robustness for the proposed simulations. To a lesser extent---in line with the observations made for both the mean and the fluctuating velocity fields---the discretization of the pressure term also seems to have an impact, especially at high Mach numbers, and primarily concentrated in the buffer-layer region. Again, given the specific features of the compared formulations, we are able to discern the impact of the discretization of the pressure terms and the fulfillment of the discrete EC property by convection, where the second one seems to be the main factor concerning overall robustness of the configuration, as shown by smaller discrepancies between the EC-TP method and the one by Gouasmi et al.~in Fig.~\ref{fig:thermo}. Also, as one could expect, the scheme by Ranocha stands in between the EC and the non-EC formulations, since, in the CP framework, it reproduces more structural properties with respect to the KEEP formulation.
\begin{table}
\centering
\normalsize
\begin{tabular}{l c c c c c c c c}
\toprule
     & $\mathrm{c}_f \times 10^3$ & $-B_q$ & $M_\tau$ & $p'^+_{w}$& $\mathrm{r}(\tilde{u}^+)$& $\mathrm{r}(\tilde{u}^+_{TL})$& $\mathrm{r}({\tau}_{11}^+)$\\
\midrule
\textbf{SCM3}  & & & &  & &\\
EC-TP &  6.78 & 0.111 & 0.126 & 1.97& -- & -- & --  \\
KEEP & 6.79 & 0.111 & 0.126   & 2.89&0.23\%&0.23\%&0.60\%\\
Gouasmi et al. & 6.84 & 0.111 & 0.127 & 1.97&0.59\%&0.57\%&1.38\%\\
\midrule
\textbf{SCM4}  & & & &  &\\
 EC-TP  &  6.79  & 0.172 & 0.147 & 2.33& --& -- & --\\
 KEEP &  6.67 & 0.171 & 0.145 & 3.23&1.16\%&1.05\%&1.59\% \\
 Gouasmi et al. &  6.81 & 0.173 & 0.147&2.31&0.23\%&0.23\%&1.21\% \\
\midrule
\textbf{HCM5}  & & & &  &\\
 EC-TP&  6.96  & 0.241 & 0.164 & 2.57&--&--&--\\
 KEEP &  6.71 & 0.237 & 0.161 & 3.55&2.08\%&1.98\%&10.98\% \\
 Gouasmi et al. &  6.89 & 0.240 & 0.163 & 2.55&0.82\%&0.78\%&5.25\% \\
 Ranocha  &  6.97 & 0.241 & 0.164 & 2.78&0.61\%&0.55\%&1.32\% \\

\bottomrule
\addlinespace
\end{tabular}
\caption{Summary of the main flow statistics: wall friction coefficient $c_f$, wall heat flux coefficient $B_q$, friction Mach number $M_\tau$, inner-scaled wall pressure root-mean-square $p'_w$ for different Mach numbers and numerical schemes. The last columns reports the relative errors with respect to the EC-TP formulation, defined as $\mathrm{r}(\phi)=\|\phi^{\mathrm{EC-TP}}-\phi\|_\infty/\|\phi^{\mathrm{EC-TP}}\|_\infty \times100$.}
\label{tab:output}
\end{table}

Table~\ref{tab:output} reports main streamwise- and spanwise-averaged flow statistics together with relative differences for the numerical discretizations under study. It is clear that non-EC schemes generally yield solutions that differ significantly from the others, especially when Mach number increases. Among all quantities, the largest discrepancies are observed for the streamwise orthogonal stress $\tau^+_{11}$ and for the skin-friction coefficient, as well as for the wall pressure root-mean-square, with the KEEP scheme being the least accurate one. Relative differences are also computed for the non-transformed streamwise velocity in viscous units, $\tilde{u}^+$, and from this analysis it seems that the TL transformation does not remarkably affect numerical differences between the considered schemes.

Finally, it is worth noting that, in practical highly compressible settings, the stabilization provided by WENO-like hybridization is often desirable. Therefore, even though the KEEP or Ranocha schemes may exhibit some discrepancies in the present tests, in practical applications they would typically be coupled with WENO-type procedures, which could potentially mitigate such issues and render them fully operational.
That said, this added robustness generally comes at the price of reduced formal cleanliness of the formulation and, to some extent, diminished solution fidelity due to the introduction of additional dissipation. In contrast, relying solely on central schemes may preserve the structural properties of the discretization more transparently and can also be advantageous from a grid-design and computational-cost perspective.
The present tests were deliberately conducted within a fully central framework, both to assess the intrinsic numerical performance of the schemes without external interference and to demonstrate that, when using EC formulations, no additional shock-capturing procedures were required for our test cases. In fact, given the smooth nature of the time-averaged flow, no strong shocks are expected. However, especially for hypersonic cases, artificial dissipation is typically introduced to counter potential instabilities. Thus, fulfilling EC property seems to deliver improvements when dealing with minor discontinuities, but in presence of strong shocks we expect a proper treatment to be still required.

\section{Conclusions}\label{sec:conclusions}
In this work, we have investigated compressible turbulent channel flows at supersonic and hypersonic Mach numbers using a thermally perfect CO$_2$ gas model, with particular emphasis on the role of structure-preserving discretizations of the convective terms in high-enthalpy regimes.
The study compares formulations with different levels of kinetic-energy preservation and entropy consistency: the widely used KEEP scheme, which is kinetic-energy-preserving and exhibits favorable entropy properties for calorically perfect gases; the Ranocha formulation, which is kinetic-energy-preserving and exactly entropy-conservative in the calorically perfect setting; the entropy-conservative scheme of Gouasmi et al., designed for thermally perfect gases but employing a pressure discretization previously shown to be suboptimal; and the EC-TP scheme, which is both exactly entropy-conservative and kinetic-energy-preserving for thermally perfect gases.

This comparison allows us to disentangle the impact of entropy consistency and pressure-term treatment when the thermodynamic closure departs from the calorically perfect assumption. The results demonstrate that entropy consistency with the underlying equation of state is a decisive factor for long-time robustness. The KEEP formulation, although reliable in calorically perfect simulations, develops significant stability issues under thermally perfect conditions. A similar but less severe trend is observed for the Ranocha scheme. In both cases, deviations first manifest in the fluctuations of thermodynamic quantities and become increasingly pronounced as the Mach number increases, eventually affecting overall robustness.

In contrast, formulations that are exactly entropy-conservative with respect to the thermally perfect model recover statistically stationary solutions across all regimes considered. However, quantitative differences remain. The Gouasmi scheme, while entropy-consistent in the convective terms, systematically overestimates Reynolds stresses, in line with its previously observed behaviors in inviscid compressible turbulence. This indicates that entropy conservation by convection, though necessary, is not alone sufficient to ensure optimal accuracy, and that the treatment of the pressure contribution influences turbulence statistics.
Among all tested discretizations, the EC-TP scheme consistently delivers the most reliable and robust results across the entire range of Mach numbers considered.
The improved performance confirms that both entropy consistency and appropriate thermodynamic coupling in the pressure treatment are essential ingredients for stable and accurate simulations of thermally perfect gases.
As expected when modifying the thermodynamic closure relations, the most pronounced discrepancies among schemes are observed in the fluctuations of thermodynamic quantities.
However, as the Mach number increases, these differences progressively influence the dynamic variables as well, highlighting the strong coupling between thermodynamics and flow dynamics in high-speed regimes.
To the best of our knowledge, the present study constitutes the first high-enthalpy channel-flow simulations of a thermally perfect gas employing exactly entropy-conservative convective schemes. The results show that \emph{ad hoc} structure-preserving formulations are not merely beneficial but necessary when addressing extreme physical conditions.

Future work will focus on extending structure-preserving concepts to the viscous terms, targeting fully entropy-stable formulations. Further developments will include applications to multi-species gas mixtures and to additional canonical configurations, such as compressible boundary layers, where thermodynamic--dynamic coupling is expected to be even more pronounced.

\clearpage

\subsection*{Acknowledgments}
We acknowledge the CINECA award under the ISCRA initiative, for the availability of high-performance computing resources and support.
\subsection*{Author Contributions}
A.~A.: Conceptualization, Methodology, Investigation, Writing – original draft, Writing – review \& editing.
A.~P.: Conceptualization, Methodology, Investigation, Writing – original draft, Writing – review \& editing.
C.~D.~M.: Conceptualization, Methodology, Investigation, Writing – original draft, Writing – review \& editing.
G.~C.: Conceptualization, Methodology, Investigation, Writing – original draft, Writing – review \& editing, Supervision.
\subsection*{Funding}
 A.P. acknowledges the financial support from ICSC - ``Centro Nazionale di Ricerca in High Performance Computing, Big Data and Quantum Computing'' funded by European Union, NextGenerationEU.
 \subsection*{Data Availability}
Data can be made available upon request to the corresponding author.


\bibliographystyle{abbrvnat}
\bibliography{refs} 
\end{document}